\documentclass[aps,prx,twocolumn,superscriptaddress,longbibliography]{revtex4-2}

\usepackage{subcaption}

\usepackage{ragged2e} % in preamble
\usepackage{braket} % for \ket and \bra commands
\usepackage{hyperref}% add hypertext capabilities

\usepackage[english]{babel}   % needed if your .bbl uses \selectlanguage

\usepackage{graphicx}% Include figure files
\usepackage{dcolumn}% Align table columns on decimal point
\usepackage{bm}% bold math

\begin{document}

\preprint{APS/123-QED}

\title{Spin-Network Quantum Reservoir Computing with Distributed Inputs: \\The Role of Entanglement}% Force 
\author{Sareh Askari}
\email{sareh.askari@ucalgary.ca}
\affiliation{Department of Physics and Astronomy, \textit{University of Calgary}, Calgary, Alberta, Canada}
\affiliation{Institute for Quantum Science and Technology (IQST), \textit{University of Calgary}, Calgary, Alberta, Canada}
\affiliation{Hotchkiss Brain Institute (HBI), \textit{University of Calgary}, Calgary, Alberta, Canada}
\author{Youssef Kora}
\affiliation{Department of Physics and Astronomy, \textit{University of Calgary}, Calgary, Alberta, Canada}
\affiliation{Institute for Quantum Science and Technology (IQST), \textit{University of Calgary}, Calgary, Alberta, Canada}
\affiliation{Hotchkiss Brain Institute (HBI), \textit{University of Calgary}, Calgary, Alberta, Canada}
\affiliation{Department of Physics, \textit{University of New Brunswick}, Fredericton, Canada}
\author{Christoph Simon}
\email{christoph.simon@ucalgary.ca}
\affiliation{Department of Physics and Astronomy, \textit{University of Calgary}, Calgary, Alberta, Canada}
\affiliation{Institute for Quantum Science and Technology (IQST), \textit{University of Calgary}, Calgary, Alberta, Canada}
\affiliation{Hotchkiss Brain Institute (HBI), \textit{University of Calgary}, Calgary, Alberta, Canada}

\date{\today}

%%%%%% Abstract %%%%%%
\begin{abstract}
Reservoir computing is a promising neuromorphic paradigm, and its quantum implementation using spin networks has shown some advantage when entanglement is present. Here, we consider a distributed scenario in which two distinct input time series are injected into separate qubits of a spin-network reservoir. We investigate how the overall entanglement, as well as its localization in the system, influence the performance of the reservoir. Focusing on bilinear memory tasks that require computing the product of the two inputs, we evaluate the short term memory capacity and correlate it with logarithmic negativity as a measure of bipartite entanglement. We find that short term memory capacity reaches its maximum at relatively small coupling strengths. In contrast, average entanglement peaks at larger couplings. Analyzing entanglement across all bipartitions, we find that the entanglement between the two input qubits is consistently the strongest and most relevant for task performance. In the small coupling strength regime where the short term memory capacity is maximized, the reservoir exhibits an extended memory tail: performance remains high for a long time. Finally, a pronounced dip in performance at zero time delay, observed across frequencies, indicates that information requires a finite propagation time through the reservoir before it can be effectively recalled. In summary, our results show that moderate entanglement—particularly between the two input qubits—plays a key role in enhancing short term memory performance.
\end{abstract}

\maketitle

%\tableofcontents

\section{Introduction}
Inspired by biological systems, neuromorphic approaches aim to mimic brain-like information processing, achieving substantial reductions in energy consumption without major compromises in performance\cite{mutlu2019processingdatamakessense, li2024neuromorphic, chen2025emerging, park2025brain, 9053043,bhatnagar2025comprehensive,cui2025bioinspired}. Among these approaches, reservoir computing (RC) stands out for its simplicity and physical realizability. The concept of reservoir computing originated in the early 2000s with the development of Echo State Networks (ESNs)\cite{jaeger2001echo} and Liquid State Machines (LSMs) \cite{maass2002real,nakajima2024review}. Unlike traditional neural networks, RC does not require training within the hidden layers; only the output layer weights are optimized \cite{374553,vrugt2024introductionreservoircomputing}. This allows RC to be implemented directly on physical systems without tuning internal parameters. An additional advantage of RC is its model-free, data-driven nature—it can infer the underlying dynamics of a system directly from time series data, without relying on an explicit model \cite{PhysRevE.111.035303, Wringe_2025}.

Building on these foundations, quantum reservoir computing (QRC) was introduced in 2017, proposing a quantum reservoir processor capable of emulating nonlinear dynamical systems—including classical chaotic behavior—through complex quantum dynamics \cite{Fujii_2017}. Subsequently, Ghosh et al. presented a quantum reservoir processing platform designed to perform quantum tasks on quantum inputs \cite{ghosh2019quantum}. Further work demonstrated that even small, noisy quantum systems can successfully perform nonlinear temporal information-processing tasks without error-mitigation techniques, marking the first experimental realization of QRC on near-term gate-model hardware \cite{PhysRevApplied.14.024065}.

Subsequent research has demonstrated several potential advantages of QRC \cite{gyurik2025quantumfeaturemapsquantum}. Notably, tasks such as prediction and reconstruction that would require thousands of neurons in classical reservoirs can be performed by only a few entangled qubits in a quantum reservoir \cite{PhysRevResearch.4.033176, kora2025statistical}. QRC has been explored on various physical platforms, including spin networks \cite{kora2024frequencydissipationdependententanglementadvantage,PhysRevA.108.052427,settino2024memory}, Rydberg atom arrays \cite{kornjača2024largescalequantumreservoirlearning,PRXQuantum.3.030325,settino2024memory, beaulieu2025robust}, Bose–Hubbard model\cite{llodra2025quantum} and nonlinear oscillators \cite{PhysRevResearch.3.013077, karimi2025role}, each offering distinct trade-offs in terms of scalability, tunability, and interaction dynamics.

It has been shown that an entanglement advantage can arise in an Ising spin network when a single time series is injected into one qubit of a four-qubit system. Specifically, they showed that the presence of entanglement enhances short-term memory when the dissipation timescale exceeds that of the input signal, highlighting its beneficial role in the system’s information-processing performance \cite{kora2024frequencydissipationdependententanglementadvantage}.

While previous work has focused on single-input scenarios, the dynamics of multi-input quantum reservoirs—especially how they depend on entanglement—remain underexplored. Recent studies have begun to address multivariate time-series settings,~\cite{hamhoum2025multivariatetimeseriesforecasting,li2025quantumreservoircomputingrealized}. However, these works primarily focus on forecasting performance and hardware feasibility, without analyzing how entanglement within the reservoir contributes to computational capability. On the other hand, entanglement provides advantages for distributed quantum information processing, particularly in communication-complexity tasks~\cite{gilboa2024exponentialquantumcommunicationadvantage, ho2022entanglement}. These results suggest that quantum correlations can enhance the exchange and integration of information between subsystems~\cite{hasegawa2025maximumseparationquantumcommunication}, providing a broader motivation for exploring multi-input quantum reservoirs.
 A recent study introduced a gate-based multivariate QRC (MTS-QRC) that assigns \(d\) \textit{injection} qubits (one per feature) and evolves them under a Hamiltonian; the approach was validated on IBM's Heron-R2 NISQ processor~\cite{hamhoum2025multivariatetimeseriesforecasting}. Another recent related study applied a similar gate-based QRC framework to real-world financial data, using a quantum reservoir to forecast a real-world financial time-series data index~\cite{li2025quantumreservoircomputingrealized}. The approach was presented as a proof-of-concept for leveraging quantum reservoirs in finance and econometrics.

In this work, we investigate the case of two input channels by injecting two distinct time series simultaneously into two separate qubits of a four-qubit Ising spin network, and we study the effect of entanglement and its structure on the short-term memory (STM) capacity \cite{jaeger2002shortterm}, denoted as \(C_{\mathrm{STM}}\). The system is governed by a Hamiltonian with randomly sampled coupling strengths, drawn from a uniform distribution. The width of this distribution serves as a hyperparameter that allows us to vary the amount of entanglement in the reservoir. Our focus is on bilinear memory tasks, in which the reservoir is trained to compute the product of the two input signals at various time delays $\tau$. These tasks are bilinear and require the reservoir to capture correlations between the two input streams. 

 Our results indicate that \(C_{\mathrm{STM}}\), increases as the average entanglement across 
all bipartitions of the system grows from zero. However, the maximum 
\(C_{\mathrm{STM}}\) is reached at relatively small coupling strengths, 
where the entanglement is only moderate and has not yet peaked. When analyzing different bipartitions of the system, we find that the two input qubits are more strongly entangled with each other than with the rest of the network, or compared to the entanglement between non-input qubits. Importantly, reservoir performance is optimal in the regime where this input–input entanglement dominates.

Moreover, we find that in this regime, the memory profile as a function of time delay 
\(\tau\) differs from other regimes. The memory decays more slowly, consistent with information being partially localized between the two input qubits due to their stronger mutual entanglement.
We also observe a pronounced dip at zero time delay in the memory profile, suggesting that information requires a finite time to propagate through the network 
before it can be effectively recalled.

The remainder of this paper is organized as follows. Section 2 outlines the methodology, including the simulation of open-system dynamics, evaluation of entanglement and memory capacity, and the definition of computational tasks. Section 3 presents the numerical results, highlighting how memory capacity depends on coupling strength and entanglement. Finally, Section 4 discusses the broader implications, emphasizing the role of input–input entanglement and contrasting weak and strong coupling regimes.
\begin{figure*}[p] % 'p' = float page, forces figure to its own page

  \centering
  % ----------- PAGE 1: Image only -----------
  \begin{subfigure}{0.9\linewidth}
    \includegraphics[width=\linewidth,clip,trim=2mm 2mm 0mm 70mm]{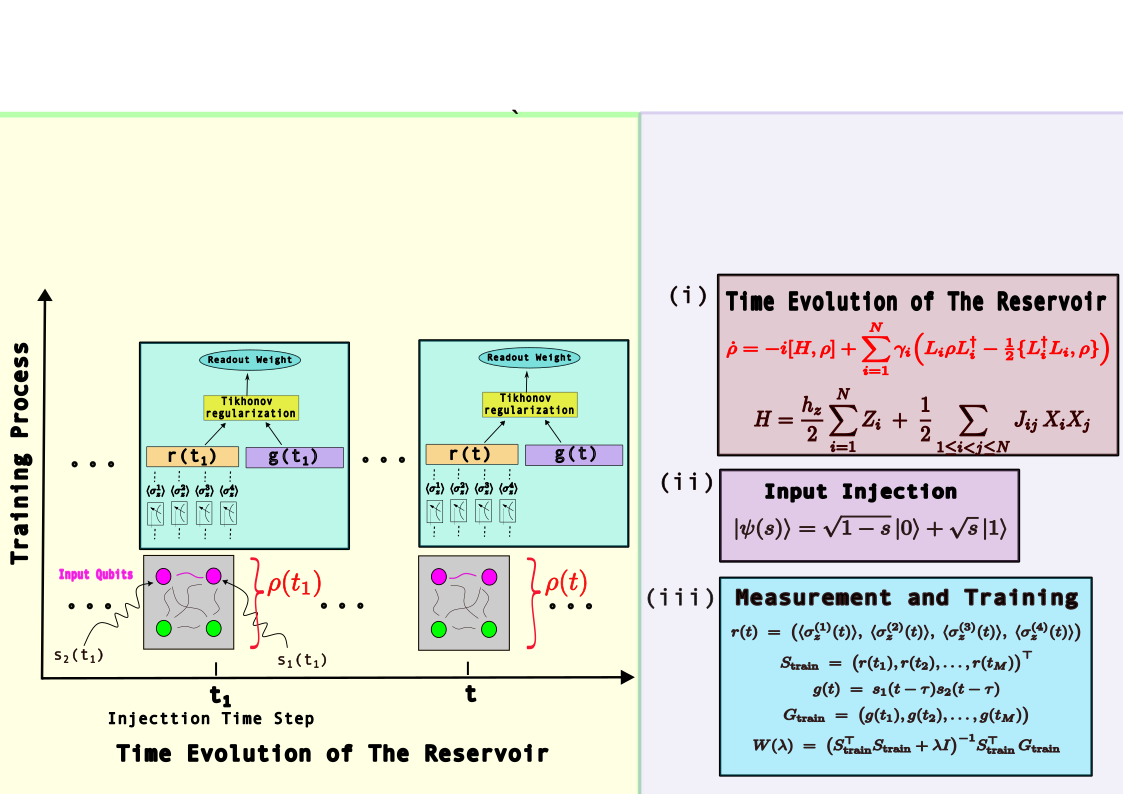}
    \caption{}
  \end{subfigure}

  \vspace{3mm}

  \begin{subfigure}{0.9\linewidth}
    \includegraphics[width=\linewidth,clip,trim=2mm 27mm 2mm 0mm]{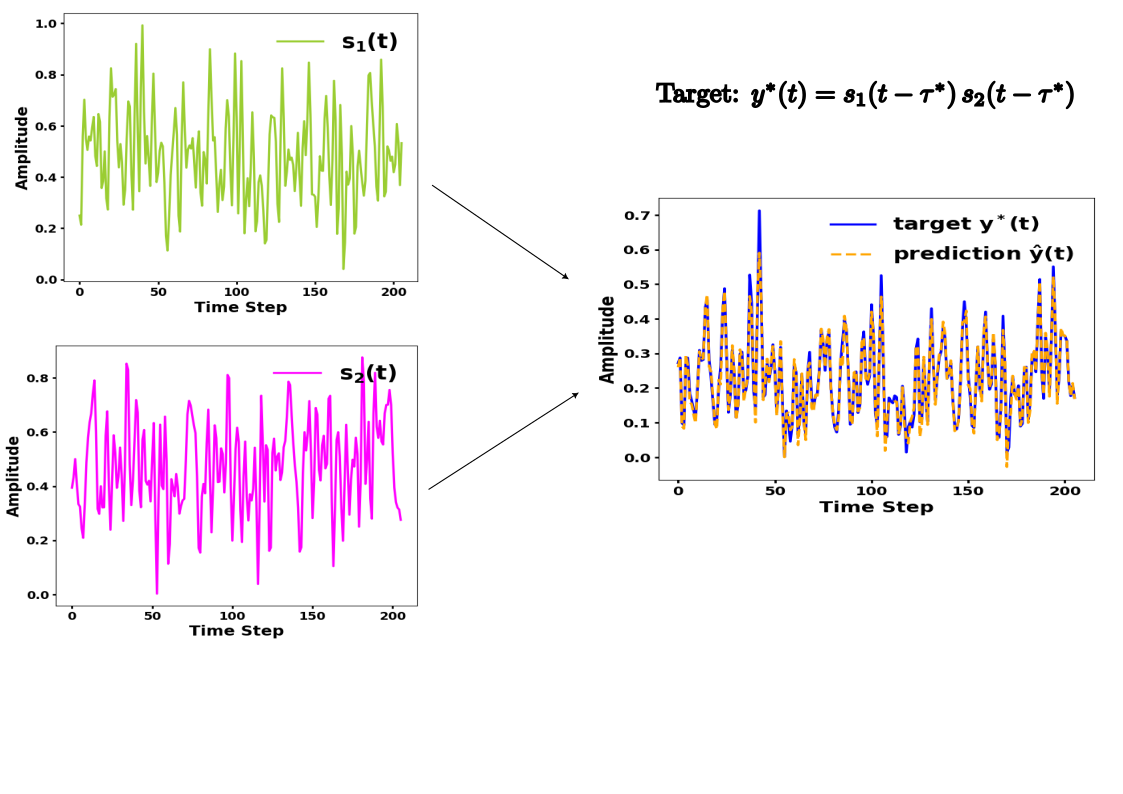}
    \caption{}
  \end{subfigure}

  \captionsetup{width=\linewidth,font=small,labelfont=bf}
  \caption{\justifying
  \textit{Distributed quantum reservoir computing pipeline and task performance.}
  \textbf{(a)} Along the horizontal axis, the reservoir---modeled as a system of four spin qubits---evolves in time under the Ising Hamiltonian [box (i)] together with environmental interactions described by the Lindblad master equation [box (i)]. At discrete injection steps \( t_k \), the two input sequences \( s_1(t_k) \) and \( s_2(t_k) \) are injected \emph{simultaneously but into separate qubits}, reflecting the distributed nature of the input encoding described in box~(ii). The reservoir then evolves from state $\rho(t_1)$ to $\rho(t)$ until the next injection, at which point the cycle repeats. The vertical line separates the physical evolution (lower part) from the learning stage (upper part). In the measurement-and-training step [box (iii)], the expectation values of all four qubits are concatenated at each time step to form a reservoir state vector $\mathbf{r}(t)$. Stacking these vectors across all times yields the training matrix $R_{\mathrm{train}}$. The target sequence, defined as $g(t)$, is collected across all time steps to form $G_{\mathrm{train}}$. This target is mapped to $R_{\mathrm{train}}$, and the readout weights $W(\lambda)$ are obtained via Tikhonov regularization.
  \textbf{(b)} Example at the optimal delay \(\tau^{\ast}\) (selected by maximizing \(C_{\mathrm{STM}}(\tau)\)) for coupling \(J_s=0.1\) and input frequency \(f=2\). Top left: input \(s_1(t)\). Bottom left: \(s_2(t)\). Top right: target \(y^{\ast}(t)=s_1(t-\tau^{\ast})\,s_2(t-\tau^{\ast})\). Bottom right: test prediction \(\hat{y}(t)\) overlaid with target \(y^{\ast}(t)\).}
  \label{fig:combined}
\end{figure*}

\section{Methodology}

Our approach involves simulating a spin-based quantum reservoir governed by a generalized transverse-field Ising Hamiltonian. The goal is to evaluate how entanglement influences memory performance in a distributed-input setting. Below, we describe the reservoir configuration, input injection method, simulation details, and performance evaluation techniques (Figure \ref{fig:combined}

% In preamble (optional if you want a real caption counter):

% In the document:

We study spin networks consisting of  4 qubits. We model the dynamics of an open quantum spin network governed by the Lindblad master equation\cite{chen2025unravelingquantumenvironmentstransformerassisted}:
\begin{equation}
\frac{d\rho}{dt} = -i[H, \rho] + \sum_i \left( L_i \rho L_i^\dagger - \frac{1}{2} \left\{ L_i^\dagger L_i, \rho \right\} \right),
\end{equation}
where $\rho$ is the density matrix of the system, $H$ is the system Hamiltonian, and $L_i$ are the Lindblad operators associated with local dissipation on each qubit.

The Hamiltonian $H$ consists of local $Z$-field terms and pairwise $XX$ interactions between qubits~\cite{MartinezPena2023}:
\begin{equation}
H = \sum_{i=1}^{4} \frac{h_z}{2} Z_i + \sum_{i<j} \frac{J_{ij}}{2} X_i X_j,
\end{equation}
where $Z_i$ and $X_i$ denote the Pauli operators acting on the $i$th qubit. We fix the local field strength to $h_z = 1.5$, which allows us to explore a broad range of ratios between the magnetic field and the interaction strength, $h_z / J_s$. This ensures that the simulations cover regimes where the dynamics are dominated by the local $Z$-field as well as those where the pairwise $XX$ couplings become the leading term.
$J_{ij}$ are the coupling coefficients between qubits $i$ and $j$. The values of $J_{ij}$ are randomly sampled from a uniform distribution in the interval $[-J_s/2, J_s/2]$, where $J_s$ is the coupling strength parameter. The coupling matrix $J_{ij}$ determines the connectivity and interaction strength within the network.

Dissipation is introduced through local Lindblad operators of the form:
\begin{equation}
L_i = \sqrt{\Gamma} \cdot \frac{1}{2}(Z_i + i Y_i),
\end{equation}
where $\Gamma=0.01$ is the dissipation rate. The chosen value is small compared to both the local field ($h_z = 1.5$) and the interaction strengths ($J_{ij} \in [-J_s/2, J_s/2]$), ensuring that dissipation acts as a weak perturbation rather than dominating the coherent dynamics. $Y_i$ is the Pauli-$Y$ operator on qubit $i$. These non-Hermitian terms model the interaction of each spin with its local environment.

We numerically integrate the evolution using a fourth-order Runge-Kutta (RK4) method \cite{Wang2023}. The time derivative $\frac{d\rho}{dt}$ is computed at each step by evaluating both the coherent unitary evolution governed by $H$ and the non-unitary dissipative dynamics from the Lindblad terms.

\subsection{Distributed Input Injection}
\paragraph{Generating input sequences}
Similar to standard practices in time–frequency benchmarking studies~\cite{miramont2024benchmarkingmulticomponentsignalprocessing}, we generate multi-component sinusoidal inputs composed of several frequency components with randomized phases to probe the reservoir’s temporal response.
We generate 9 distinct time series, each composed of 20 sinusoidal components with frequencies 
\( f_k \in \left[ \frac{f_0}{5000}, \frac{f_0}{50} \right] \), 
linearly spaced across this range. Independent phase offsets 
\( \phi_{k} \sim \mathcal{U}(0,1) \) 
are used to introduce variability across sequences and components. 
The time variable \( t \in [0,1] \) is sampled at \( N = 1000 \) discrete time steps. 
Each sequence is constructed according to:

Each input sequence is constructed as a sum of sinusoidal components,
\[
x_{q}(t) = \sum_{k=1}^{20} \sin\!\left( 2\pi f_k t + 2\pi \phi_k^{(i)} \right),
\]
and is then normalized to the range \([0,1]\) using min–max normalization:
\[
s_{q}(t) =
\frac{x_{q}(t) - \min_t \big( x_{q}(t) \big)}
     {\max_t \big( x_{q}(t)\big) - \min_t \big( x_{q}(t) \big)}.
\]
Here the subscript \( q \) indicates the qubit to which the signal is injected (e.g., \(q = 1, 2\) for input qubits 1 and 2).

% ___________________________________________________________
% To incorporate external inputs into the quantum reservoir, we periodically reset a subset of qubits to pure states encoding classical signals. Specifically, at regular time intervals $\Delta t_{\text{inj}} = n_{\text{bet}} \cdot \delta t$, we update the reduced state of the input qubits by replacing them with pure states constructed from the input time series.

% Each input state $\psi(s_k)$ is defined as:
% \begin{equation}
% \psi(s_k) = \sqrt{1 - s_k} \ket{0} + \sqrt{s_k} \ket{1},
% \end{equation}

% where $s_k \in [0,1]$ is the $k$th input value. This corresponds to a superposition of the computational basis states, encoding $s_k$ in the amplitudes of the qubit.
% To inject classical input signals into the quantum reservoir, we periodically reset two designated input qubits to pure states that encode the input values at each time step. Let $\rho$ be the full density matrix of the $N$-qubit system, and let qubits indexed by subsets $\mathcal{S}_1$ and $\mathcal{S}_2$ correspond to the input subsystems to be updated with pure states $\ket{\psi_1}$ and $\ket{\psi_2}$, respectively.
\paragraph{Input Injection.} The total evolution time was set to $t_{\mathrm{final}} = 2750$ (in dimensionless units), 
and input signals were injected periodically every $\Delta t = 7.5$ time units. 
This choice yielded the best overall memory performance across the tested coupling strengths and input frequencies, 
indicating an optimal trade-off between information retention and refresh of input signals.
We inject classical inputs $s_1$ and $s_2$ from two independent sequences by resetting two designated qubits (typically qubits 1 and 2) to pure states that encode the current input values. Each input is mapped to a single-qubit state defined as: 
\begin{equation}
\ket{\psi(s)} = \sqrt{1 - s} \ket{0} + \sqrt{s} \ket{1}
\end{equation}
where $s \in [0,1]$ is the normalized input value at the current time step.

To inject the new input states, we first remove the original states of the input qubits by computing the partial trace:
\[
\rho_{34} = \text{Tr}_{\text{12}} \left(\rho) \right.
\]
We then construct the updated density matrix by taking the tensor product of the reduced reservoir state with the new input states:
\begin{equation}
\rho' =\rho_1({\text{s}_1}) \otimes \rho_2({\text{s}_2}) \otimes\rho_{34}.
\end{equation}
where $\rho_1({\text{s}_1}) = \ket{\psi(\text{s}_1)}\bra{\psi(\text{s}_1)}$ and similarly for $\rho_2({\text{s}_2})$.

\subsection{Bilinear short-term memory (STM) task}

We evaluate the reservoir's capability to retain and process temporal information using a bilinear short-term memory (STM) task. 
At each time step \(k\), the target output is defined as the element-wise product of the two input signals with a temporal delay \(\tau\):
\begin{equation}
\bar{y}_k = s_1(k - \tau) s_2(k - \tau).
\end{equation}
This task requires the reservoir to store past information from both input streams and to reproduce their temporal correlations at a later time, thus serving as a measure of its memory capacity.

%\subsection{Reservoir Readout and Task Definition}
\paragraph{Training the reservoir} From each input sequence, we sample values at fixed intervals. These sampled input values are then used to construct the target functions for training and evaluation. The reservoir outputs—given by the expectation values \( \langle \sigma^{i}_z \rangle \) from all qubits i —are collected at each injection time, after the reservoir has reached its steady state, and assembled into a feature matrix. A linear regression model is then trained to map the reservoir outputs to the corresponding target function, with or without delay, thereby enabling the reservoir to retain and process short-term temporal information.

% To extract features for learning, we measure the expectation values of $\sigma_z$ for all qubits at each input step. These measurements are concatenated into feature vectors, which are used in a supervised learning task.
\paragraph{Testing the reservoir's output}
In the testing phase, we evaluate the generalization performance of the reservoir using two previously unseen input sequences.  As described previously, after feeding the test sequences into the reservoir and allowing it to reach a steady state, we collect the $\langle \sigma_z \rangle$ expectation values from the readout qubits and assemble them into a test feature matrix.

\paragraph{Ridge regression and the regularization parameter $\lambda$} To obtain the optimal output weights, we use ridge regression (also known as Tikhonov regularization). 
Given the training data $R$ (reservoir states) and target outputs $G$, 
the regression seeks a linear map $A$ that minimizes the regularized loss function
\begin{equation}
\mathcal{L}(A_{out}) = \| G - A_{out} R \|^2 + \lambda \| A_{out} \|^2,
\end{equation}
where $\lambda$ controls the trade-off between fitting accuracy and weight regularization.

The analytical solution is given by
\begin{equation}
A_{out} = G  R^{\mathrm{T}}
\left(R R^{\mathrm{T}} + \lambda I \right)^{-1},
\end{equation}
where $I$ is the identity matrix. A larger $\lambda$ suppresses large weight amplitudes and prevents overfitting, 
while a smaller $\lambda$ allows more faithful fitting \cite{Griffith2021}.

The performance of the model is quantified using the squared Pearson correlation coefficient between the predicted and true outputs. To evaluate the overall memory capacity of the reservoir, we define the short-term memory (STM) capacity at each $\tau$ as\cite{article}:

\begin{equation}
C_{\mathrm{STM}}(\tau) = \frac{\mathrm{cov}^2(y, \bar{y}_{\tau})}{\sigma_y^2 \, \sigma_{\bar{y}_{\tau}}^2},
\end{equation}

where \( y \) is the predicted output vector, \( \bar{y}_{\tau} \) is the delayed target sequence, \( \mathrm{cov}(y, \bar{y}_{\tau}) \) denotes their covariance, and \( \sigma \) represents the standard deviation. The optimal regularization parameter \( \lambda \) for ridge regression is selected by maximizing the test-time memory performance over a logarithmically spaced range of values. This ensures robust generalization while avoiding overfitting.
To quantify the reservoir’s memory performance, we compute the total short-term memory capacity as
\begin{equation}
C_{\mathrm{STM}} = \sum_{\tau=0}^{\infty} C_{\mathrm{STM}}(\tau).
\end{equation}
In practice, the summation is truncated once $C_{\mathrm{STM}}{(\tau)}$ becomes negligible. $C_{\mathrm{STM}}{(\tau)}$ has already decayed to zero by $\tau = 24$ in all cases considered here, 
so we restrict the calculation to this range.

% We consider a bilinear short-term memory (STM) task, where the target output at each timestep is:

% \begin{equation}
% \bar{y}k = s{k - \tau}^{(1)} \cdot s_{k - \tau}^{(2)},
% \end{equation}

% with $\tau$ being the time delay. This task requires the reservoir to retain and correlate past values of the two input streams. A linear regression model is trained to map reservoir outputs to the target sequence, and performance is quantified using the squared Pearson correlation coefficient between predicted and true outputs.

% To evaluate overall memory performance, we define the total short-term memory capacity as:

% \begin{equation}
% C_{\text{STM}} = \sum_{\tau} \frac{\text{cov}^2(y, \bar{y}\tau)}{\sigma_y^2 \sigma{\bar{y}_\tau}^2},
% \end{equation}

% where $y$ is the predicted output vector, $\bar{y}\tau$ is the delayed target sequence, $\text{cov}(y, \bar{y}\tau)$ is their covariance, and $\sigma$ denotes the standard deviation.

% In addition, we define a weighted memory capacity that emphasizes instantaneous memory:

% \begin{equation}
% C_{\text{STM}}^{\text{(weighted)}} = \sum_{\tau = 0}^{25} \frac{\text{cov}^2(y, \bar{y}\tau)}{\sigma_y^2 \sigma{\bar{y}_\tau}^2} + 25 \cdot \frac{\text{cov}^2(y, \bar{y}0)}{\sigma_y^2 \sigma{\bar{y}_0}^2},
% \end{equation}

% giving 26 times more weight to the memory at $\tau = 0$ compared to other delays. This metric is designed to capture the reservoir’s ability to respond immediately to new input correlations—an important consideration when information has not yet propagated through the network.
\begin{figure*}[p]
  \centering

  % --- Left: image (top-aligned) ---
  \begin{minipage}[t]{0.5\textwidth}
    \vspace{0pt} % <-- forces the first baseline to be at the top
    
    \includegraphics[width=\linewidth,keepaspectratio,clip,trim=0 8 0 0]{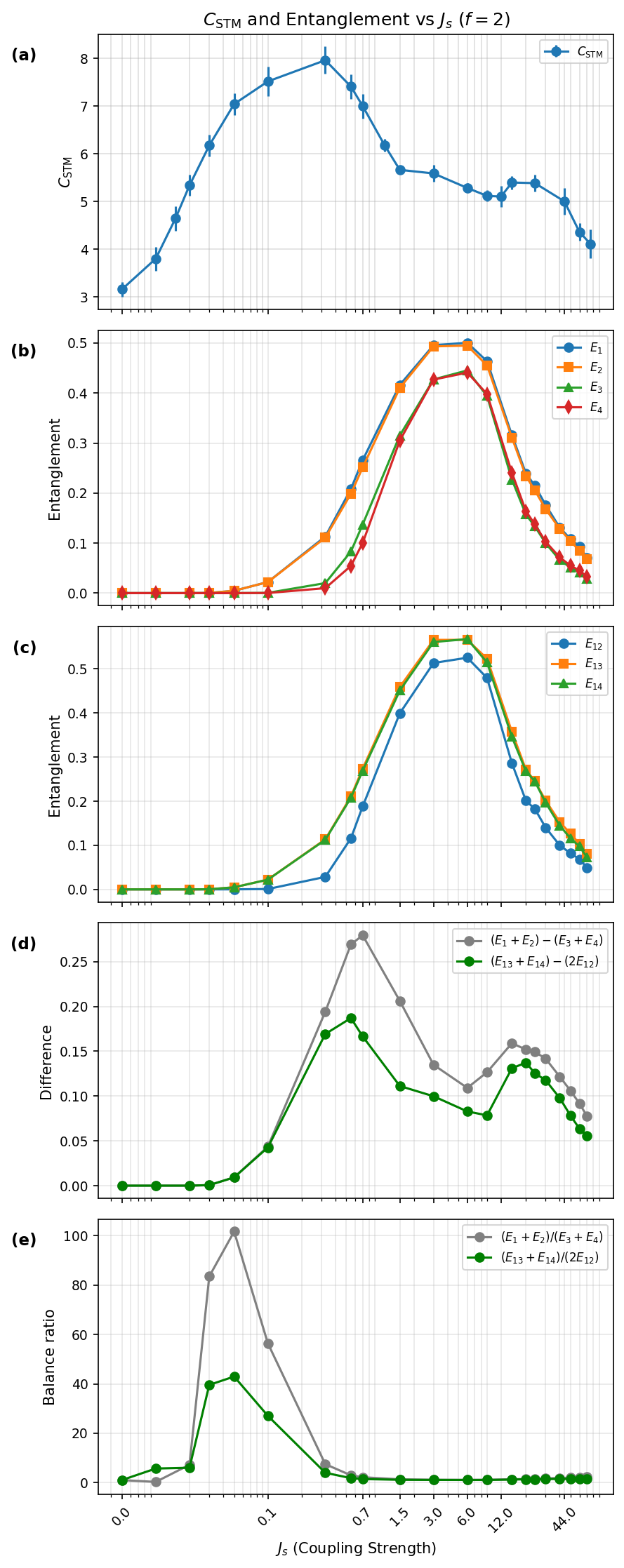}
    % ^ trim top whitespace if your PDF has margins; adjust the '8' if needed
    
     \label{fig:c_ent_J_s}
  \end{minipage}\hfill
  % --- Right: caption (top-aligned) ---
  \begin{minipage}[t]{0.42\textwidth}
    \vspace{0pt} % <-- align the top line with the image top
    \small
    % If you want a REAL caption (numbered, list-of-figures):
    \setlength{\abovecaptionskip}{0pt} % <-- remove space above caption text
    \captionsetup{width=\linewidth,font=small,labelfont=bf}
    \captionof{figure}{\justifying
        \textit{Memory capacity, average entanglement, and input–input entanglement in different coupling regimes.}
        \textbf{(\textsf{a)}} Memory capacity vs.\(J_s\) peaks at intermediate couplings (\(J_s \approx 0.3\)) before decaying.
        \textbf{\textsf{(b)}} Curves show the logarithmic negativity \(E\), averaged over ten random connectivity matrices.
        Entanglements correspond to different partitions of the system:
        \(E_1:{\{1\}\,|\,\{2,3,4\}},
        \,E_2:{\{2\}\,|\,\{1,3,4\}},
        \,E_3:{\{3\}\,|\,\{1,2,4\}},
        \,E_4:{\{4\}\,|\,\{1,2,3\}}\).
        (\(E_1\)–\(E_4\)) show similar non-monotonic behavior with a maximum near \(J_s \approx 6\).
        \textbf{(\textsf{(c)}} Pair–vs.–rest partitions are
        \(E_{12}:{\{1,2\}\,|\,\{3,4\}},
        \,E_{13}:{\{1,3\}\,|\,\{2,4\}},\,
        E_{14}:{\{1,4\}\,|\,\{2,3\}}\),
        following the same trend with modest differences.
        At \(f=2\), larger \(E_1\) and \(E_2\) together with \(E_{12}<E_{13},E_{14}\) indicate that entanglement is concentrated within the input pair \(\{1,2\}\).
        \textbf{(\textsf{(d)}} Gray curve: difference between $E_1+E_2$ and $E_3+E_4$; green curve: difference between $E_{13}+E_{14}$ and $2E_{12}$. Both exhibit maxima around $J_s \approx 0.3$--$0.7$, following a trend similar to panel~(a). 
        \textbf{(\textsf{(e)}} Balance ratios: $(E_1+E_2)/(E_3+E_4)$ and $(E_{13}+E_{14})/(2E_{12})$, both peaking around $J_s \approx 0.05$. 
        Overall, the optimum performance occurs along the initial rise before the entanglement peak, where the entanglement between qubits 1 and 2 dominates relative to other partitions—suggesting that this form of entanglement is the most beneficial for the task.}
  \label{fig:c_ent_J_s}

    % If you prefer manual caption text instead of captionof, comment the 3 lines above
    % and start your text directly (no extra vertical skip will be inserted).
  \end{minipage}
\end{figure*}

\subsection{Entanglement Analysis}

To quantify bipartite entanglement within the reservoir, we use the logarithmic negativity. 
For a bipartition between subsystem \(A\) and the remainder of the system, it is defined as
\begin{equation}
E_{A}(\rho) = \log \left\| \rho^{T_A} \right\|_1,
\end{equation}
where \(\rho^{T_A}\) denotes the partial transpose of the density matrix with respect to subsystem \(A\) and logarithm is taken in base 2. Here, \(A\) can correspond either to a single qubit \(i\)
or to a pair of qubits \(ij\). We denote by \(E_i\) the entanglement between qubit \(i\) and the rest of the system, 
and by \(E_{ij}\) the entanglement between the pair of qubits \(\{i,j\}\) and the remaining qubits, 
both corresponding to the general definition \(E_A(\rho)\) for subsystem \(A\). All logarithmic negativities are computed from the steady-state density matrix $\rho$.
During the simulation, we evolve the full density matrix \( \rho(t) \) of the open quantum system under Ising hamiltonian and Lindblad dynamics and extract it at regular intervals. At each sampled time step, we compute the logarithmic negativity for multiple bipartitions of the four-qubit network. 
At each time step, the corresponding entanglement values are denoted by \( {E}_i(t) \) or \( {E}_{ij}(t) \) for single–vs.–rest and pair–vs.–rest partitions, respectively. These quantities are then averaged over time after the reservoir has reached its steady state.

%For a global characterization of entanglement across the entire reservoir, we define the average (or total) entanglement as
%\[
%\mathrm{Ent} = \frac{1}{7} \left( \overline{{E}1} + \overline{{E}2} + \overline{{E}3} + \overline{{E}4}
% +\overline{{E}{12}} + \overline{{E}{13}} + \overline{{E}{14}} \right).
%\]

This quantity provides a single scalar metric for tracking how entanglement is distributed throughout the network and how it evolves with coupling strength \(J_s\). In the results section, we focus primarily on the behavior of single–vs.–rest partitions (\(E_1\)–\(E_4\)), pair–vs.–rest partitions (\(E_{12}\), \(E_{13}\), \(E_{14}\)), to examine how different forms of entanglement correlate with memory capacity and task performance.

\section{Results}
\subsection{Memory capacity and average entanglement in different coupling regimes.}

Figure~\ref{fig:c_ent_J_s}a shows that the optimal performance is reached during the initial rise of memory capacity, 
before the peak of average entanglement, rather than at the maximum entanglement point. Specifically, 
the memory capacity increases going from \(J_s \approx 0\) to \(J_s \approx 0.3\). 
In this regime, the entanglement also grows. From \(J_s \approx 0.3\) to \(J_s = 6\), the memory capacity decreases again, while 
entanglement continues to grow, reaching its maximum value. Beyond this point, 
further increases in the coupling strength reduce both entanglement and memory capacity.

Figure~2b reveals that \(E_1\) and \(E_2\) are consistently larger than \(E_3\) and \(E_4\).

Figure~2c further shows that $E_{12}$ is larger than $E_{13}$ and $E_{14}$. Combined with the observations from Fig.~2b, this establishes 
that the entanglement between qubits~1 and~2 dominates over other types of entanglement in the system.

When examining Fig.~\ref{fig:c_ent_J_s}{a–c}, it appears that the memory performance peaks around \( J_s \approx 0.3 \), coinciding with regions where the entanglement associated with the input qubits (\(E_1, E_2\)) differs most strongly from that of the remaining qubits (\(E_3, E_4\)), and where the pairwise input entanglements (\(E_{13}, E_{14}\)) become more pronounced relative to \(E_{12}\). 
To quantify this observation, we compute the following differences and ratios of entanglement measures.
The differences 
\((E_1+E_2)-(E_3+E_4)\) and \((E_{13}+E_{14})-(2E_{12})\) 
both reach their maxima at $J_s \approx 0.5$, which is near  the coupling strength corresponding to the peak of \( C_{\mathrm{STM}} \) ($J_s \approx0.32$), whereas the overall entanglement peaks at \( J_s \approx 6 \) (where the performance is not maximum). The trends of these differences 
closely follow that of \( C_{\mathrm{STM}} \) versus \( J_s \), particularly 
for larger couplings (\( J_s > 6 \)) (Fig.~\ref{fig:c_ent_J_s}d).  

% The normalized entanglement ratios,  
% \[
% \frac{E_1+E_2}{E_1+E_2+E_3+E_4} \quad \text{and} \quad \frac{E_{13}+E_{14}}{E_{13}+E_{14}+E_{12}},
% \]
% show a maximum at \( J_s \approx 0.1 \), which is close to the coupling strength where \( C_{\mathrm{STM}} \) attains its maximum.

Similarly, the balance ratios,  
\[
\frac{E_1+E_2}{E_3+E_4} \quad \text{and} \quad \frac{E_{13}+E_{14}}{2E_{12}},
\]
exhibit a maximum at approximately the same coupling strength—again close to the point of highest \( C_{\mathrm{STM}} \) (Fig.~\ref{fig:c_ent_J_s}e).
  \begin{figure*}[t]
    \centering
    \begin{subfigure}{0.48\textwidth}
        \centering
        \includegraphics[width=\linewidth]{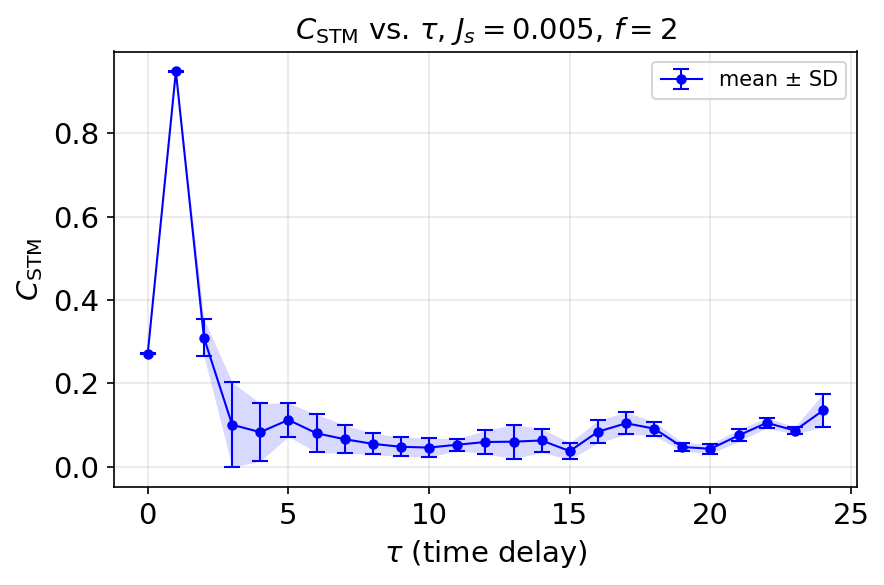}
        \caption{$J_s=0.005$}
    \end{subfigure}\hfill
    \begin{subfigure}{0.48\textwidth}
        \centering
        \includegraphics[width=\linewidth]{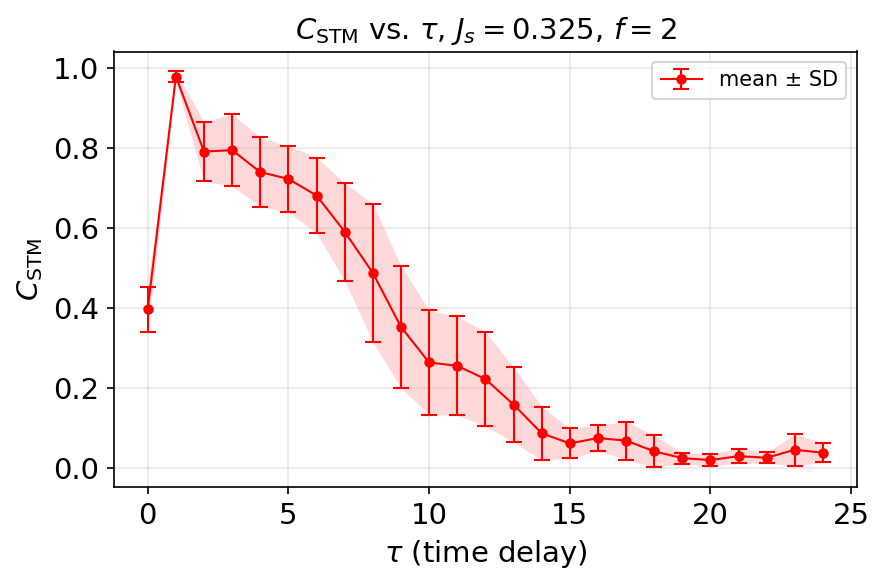}
        \caption{$J_s=0.325$}
    \end{subfigure}\hfill 
        % Add vertical space between rows

    \vspace{1em} % <-- adjust this value

    \begin{subfigure}{0.48\textwidth}
        \centering
        \includegraphics[width=\linewidth]{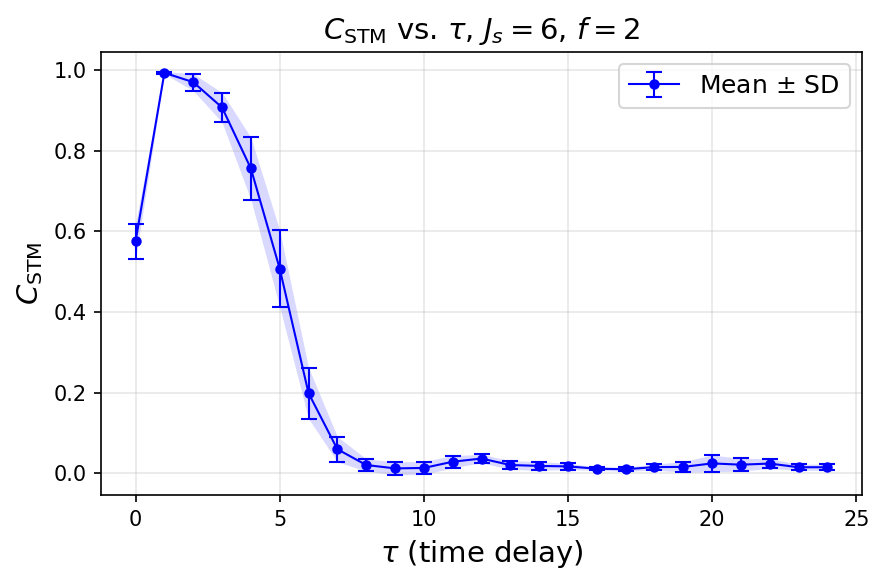}
        \caption{$J_s=6$}
    \end{subfigure}\hfill
    \begin{subfigure}{0.48\textwidth}
        \centering
        \includegraphics[width=\linewidth]{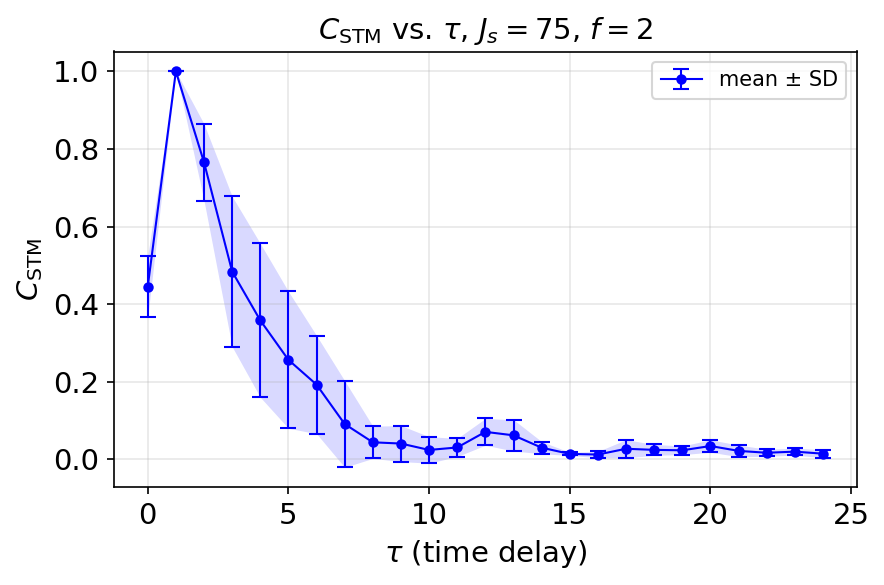}
        \caption{$J_s=75$}
    \end{subfigure}
    \captionsetup{width=\linewidth,font=small,labelfont=bf}
    \caption{\justifying Short-term memory capacity $C_{\mathrm{STM}}(\tau)$ versus delay $\tau$ at three coupling strengths. 
    At very small coupling ($J_s=0.005$), the memory profile resembles that of the moderate and high-coupling case ($J_s=6$ and $J_s=75$), with a fast decaying memory. By contrast, the weak-coupling case ($J_s=0.325$) exhibits a long, low-amplitude tail (slow fading memory). 
    These distinct decay behaviors correlates with Fig.~\ref{fig:c_ent_J_s}: in figure Fig.~\ref{fig:c_ent_J_s}, the performance peaks at $J_s \approx0.3$, here we can see that at this regime the memory profile is not similar to other regimes, exhibiting a slower decay. The difference can be understood as information becoming trapped between the two input qubits at $J_s=0.325$, consistent with Fig.~\ref{fig:c_ent_J_s}, which shows that the input qubits are more strongly entangled with each other than with the rest of the network.}
    \label{fig:CstmTau_compare}
\end{figure*}
All these quantities reach their maxima within approximately the same coupling regime— an order of magnitude weaker than the coupling strength at which the entanglements attain their peak. Taken together, these results indicate that memory capacity peaks when the 
entanglement between the two input qubits (1 and 2) dominates over other 
partitions, suggesting that this particular form of entanglement is especially 
relevant for the computational task.

\begin{figure*}[!t]
    \centering
    \begin{subfigure}{0.32\textwidth}
        \centering
        \includegraphics[width=\linewidth]{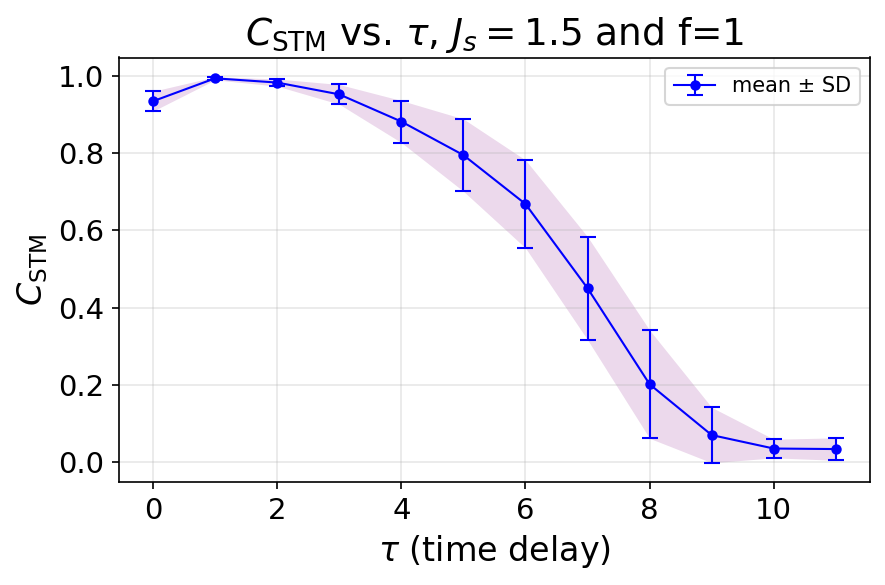}
        \caption{$f=1$}
    \end{subfigure}\hfill
    \begin{subfigure}{0.32\textwidth}
        \centering
        \includegraphics[width=\linewidth]{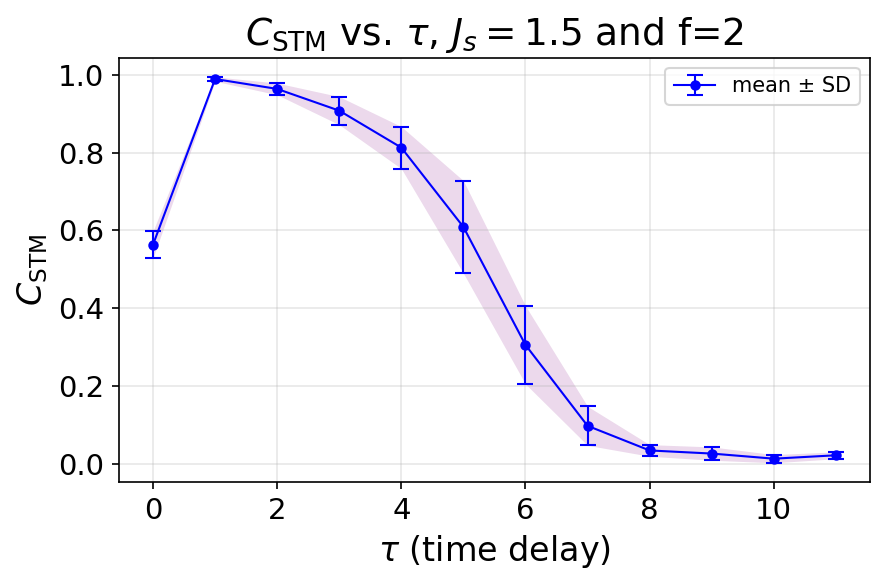}
        \caption{$f=2$}
    \end{subfigure}\hfill
    \begin{subfigure}{0.32\textwidth}
        \centering
        \includegraphics[width=\linewidth]{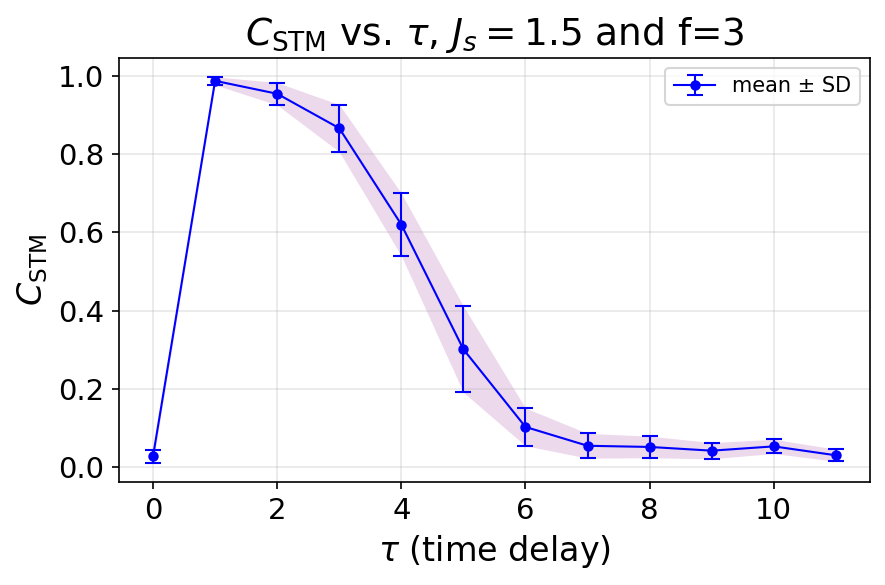}
        \caption{$f=3$}
    \end{subfigure}
    \captionsetup{width=\linewidth,font=small,labelfont=bf}
    \caption{\justifying
    Short-term memory capacity $C_{\mathrm{STM}}(\tau)$ versus delay $\tau$ at fixed coupling $J_s=1.5$ for three driving frequencies ($f\in\{1,2,3\}$). 
   All curves exhibit a dip at \(\tau = 0\), a near-term peak, and a subsequent decay. 
   The depth of the zero-delay dip increases with \(f\), consistent with an increasingly challenging task at higher driving frequencies. 
This initial dip reflects the fact that, in our distributed setting, information requires additional time to propagate throughout the reservoir before it can be effectively recalled.}

    \label{fig:CstmTau_J_s15}
\end{figure*} 

\subsection{Memory tail in different coupling regimes}

Examining the memory capacity as a function of time delay \(\tau\) across 
different coupling strengths, we find that for very small or nearly vanishing 
couplings the memory decays around \(\tau \approx 8\). (Fig.~\ref{fig:CstmTau_compare}a)
For both intermediate couplings 
(\(J_s \approx 6\)) and strong couplings (\(J_s \approx 75\)), the memory also 
decays rapidly and vanishes near \(\tau \approx 8\). (Fig.~\ref{fig:CstmTau_compare}c and
Fig.~\ref{fig:CstmTau_compare}d)

In contrast, for a lower but finite coupling of \(J_s \approx 0.3\), a 
pronounced long memory tail emerges, extending until approximately 
\(\tau \approx 18\). (Fig.~\ref{fig:CstmTau_compare}b) As discussed in Sec.~2.1, in this regime the input qubits 
are more strongly entangled with each other than with the rest of the network. 
The appearance of a long memory tail therefore suggests that information tends 
to remain localized between the input qubits, slowing the overall decay of 
memory.

\subsection{Dip in Memory profile}
A pronounced dip is consistently observed in the memory profile as a function 
of time delay \(\tau\). The memory capacity is strongly suppressed at zero 
delay, corresponding to the moment when the input is injected (Fig.~\ref{fig:CstmTau_J_s15}).
It subsequently 
rises, reaching nearly \(C_{\mathrm{STM}}(1) \approx 1\). For 
longer delays, the behavior depends on the input frequency: at \(f=1\), the 
capacity remains relatively high at \(C_{\mathrm{STM}} \approx 0.9\); for 
\(f=2\), it decreases to about \(C_{\mathrm{STM}} \approx 0.6\); and for 
\(f=3\), it vanishes entirely (\(C_{\mathrm{STM}} = 0\)). Thus, the dip at 
zero delay becomes more pronounced as the task difficulty increases (Fig.~\ref{fig:CstmTau_J_s15}).

This behavior reflects the fact that the system requires a finite amount of 
time to distribute information throughout the reservoir before it can be 
effectively recalled.

\section{Discussion}

Our results indicate that in our distributed-input setting, the best performance does not occur at the strongest coupling or at the highest level of average entanglement. Instead, it emerges in a regime characterized by moderate coupling and localized entanglement. In this optimal regime, the entanglement between the two input qubits is stronger than that involving other parts of the reservoir.
This concentration of entanglement on the input qubits appears to be particularly beneficial for computation. It may enable the reservoir to retain and process correlations from the input time series efficiently, without excessive mixing with the rest of the network. The moderate coupling ensures communication between the inputs while maintaining a degree of isolation that preserves useful correlations.
Overall, these observations suggest that it is not only the amount of entanglement but also its structure—specifically, its localization on the input nodes—that matters for short-term memory performance in distributed quantum reservoirs.

A further notable feature is the pronounced dip in memory performance at zero time delay. 
This dip, which becomes sharper at higher input frequencies, indicates that information cannot be recalled instantaneously at the time of injection. 
Instead, the reservoir requires a finite time to propagate and redistribute the input across the network before it can be effectively utilized. 

These findings open several directions for future research. It would be interesting to investigate whether similar relationships between entanglement structure and memory capacity appear in larger spin networks, especially on experimentally accessible platforms. Another natural extension would be to examine how the performance changes when different inputs are injected into more than two qubits. Another promising direction would be to explore the effect of varying the coupling strength between the input qubits—making it stronger or weaker than the other couplings in the network—and examining how such asymmetries influence different forms of entanglement and the reservoir’s overall performance. Moreover, applying the same analysis to other benchmark tasks, such as nonlinear autoregressive moving average (NARMA) sequences, could help assess the generality of the observed trends and clarify the broader computational capabilities of spin-network reservoirs. Such tasks were not explored here due to computational constraints but represent an important direction for future work.\\
Overall, examining how the amount and structure of entanglement within such network systems affect their memory capacity may help bridge quantum dynamics and information processing, paving the way toward more advanced machine-learning and neural-network algorithms that harness quantum properties for distributed computation.
\section*{Code Availability}
The simulation and analysis codes used in this study are available in 
\href{https://github.com/sarehaskari17/spin-network-qrc.git}{\texttt{GitHub}}. 
Additional scripts and data are available from the corresponding author upon reasonable request.

\section*{Acknowledgements}

This work was supported by the National Research Council of Canada through its Applied Quantum Computing challenge program, the Natural Sciences and Engineering Research Council through its Discovery Grant program, by an Alberta Innovates Discovery Grant supplement, and by Quantum City.

\nocite{*}
% \bibliographystyle{apsrev4-2}
% \bibliography{ref}

\begin{thebibliography}{43}%
\makeatletter
\providecommand \@ifxundefined [1]{%
 \@ifx{#1\undefined}
}%
\providecommand \@ifnum [1]{%
 \ifnum #1\expandafter \@firstoftwo
 \else \expandafter \@secondoftwo
 \fi
}%
\providecommand \@ifx [1]{%
 \ifx #1\expandafter \@firstoftwo
 \else \expandafter \@secondoftwo
 \fi
}%
\providecommand \natexlab [1]{#1}%
\providecommand \enquote  [1]{``#1''}%
\providecommand \bibnamefont  [1]{#1}%
\providecommand \bibfnamefont [1]{#1}%
\providecommand \citenamefont [1]{#1}%
\providecommand \href@noop [0]{\@secondoftwo}%
\providecommand \href [0]{\begingroup \@sanitize@url \@href}%
\providecommand \@href[1]{\@@startlink{#1}\@@href}%
\providecommand \@@href[1]{\endgroup#1\@@endlink}%
\providecommand \@sanitize@url [0]{\catcode `\\12\catcode `\$12\catcode
  `\&12\catcode `\#12\catcode `\^12\catcode `\_12\catcode `\%12\relax}%
\providecommand \@@startlink[1]{}%
\providecommand \@@endlink[0]{}%
\providecommand \url  [0]{\begingroup\@sanitize@url \@url }%
\providecommand \@url [1]{\endgroup\@href {#1}{\urlprefix }}%
\providecommand \urlprefix  [0]{URL }%
\providecommand \Eprint [0]{\href }%
\providecommand \doibase [0]{https://doi.org/}%
\providecommand \selectlanguage [0]{\@gobble}%
\providecommand \bibinfo  [0]{\@secondoftwo}%
\providecommand \bibfield  [0]{\@secondoftwo}%
\providecommand \translation [1]{[#1]}%
\providecommand \BibitemOpen [0]{}%
\providecommand \bibitemStop [0]{}%
\providecommand \bibitemNoStop [0]{.\EOS\space}%
\providecommand \EOS [0]{\spacefactor3000\relax}%
\providecommand \BibitemShut  [1]{\csname bibitem#1\endcsname}%
\let\auto@bib@innerbib\@empty
%</preamble>
\bibitem [{\citenamefont {Mutlu}\ \emph {et~al.}(2019)\citenamefont {Mutlu},
  \citenamefont {Ghose}, \citenamefont {Gómez-Luna},\ and\ \citenamefont
  {Ausavarungnirun}}]{mutlu2019processingdatamakessense}%
  \BibitemOpen
  \bibfield  {author} {\bibinfo {author} {\bibfnamefont {O.}~\bibnamefont
  {Mutlu}}, \bibinfo {author} {\bibfnamefont {S.}~\bibnamefont {Ghose}},
  \bibinfo {author} {\bibfnamefont {J.}~\bibnamefont {Gómez-Luna}},\ and\
  \bibinfo {author} {\bibfnamefont {R.}~\bibnamefont {Ausavarungnirun}},\
  }\href {https://arxiv.org/abs/1903.03988} {\bibinfo {title} {Processing data
  where it makes sense: Enabling in-memory computation}} (\bibinfo {year}
  {2019}),\ \Eprint {https://arxiv.org/abs/1903.03988} {arXiv:1903.03988
  [cs.AR]} \BibitemShut {NoStop}%
\bibitem [{\citenamefont {Li}\ \emph {et~al.}(2024)\citenamefont {Li},
  \citenamefont {Wen}, \citenamefont {Liu}, \citenamefont {Tan}, \citenamefont
  {Deng}, \citenamefont {Pei},\ and\ \citenamefont {Shi}}]{li2024neuromorphic}%
  \BibitemOpen
  \bibfield  {author} {\bibinfo {author} {\bibfnamefont {M.}~\bibnamefont
  {Li}}, \bibinfo {author} {\bibfnamefont {X.}~\bibnamefont {Wen}}, \bibinfo
  {author} {\bibfnamefont {X.}~\bibnamefont {Liu}}, \bibinfo {author}
  {\bibfnamefont {Y.}~\bibnamefont {Tan}}, \bibinfo {author} {\bibfnamefont
  {L.}~\bibnamefont {Deng}}, \bibinfo {author} {\bibfnamefont {J.}~\bibnamefont
  {Pei}},\ and\ \bibinfo {author} {\bibfnamefont {L.}~\bibnamefont {Shi}},\
  }\href {https://doi.org/10.1038/s41586-024-08253-8} {\bibfield  {journal}
  {\bibinfo  {journal} {Nature}\ }\textbf {\bibinfo {volume} {628}},\ \bibinfo
  {pages} {974} (\bibinfo {year} {2024})}\BibitemShut {NoStop}%
\bibitem [{\citenamefont {Chen}\ \emph {et~al.}(2025)\citenamefont {Chen},
  \citenamefont {Zhou}, \citenamefont {Tong}, \citenamefont {Pang},\ and\
  \citenamefont {Xu}}]{chen2025emerging}%
  \BibitemOpen
  \bibfield  {author} {\bibinfo {author} {\bibfnamefont {C.}~\bibnamefont
  {Chen}}, \bibinfo {author} {\bibfnamefont {Y.}~\bibnamefont {Zhou}}, \bibinfo
  {author} {\bibfnamefont {L.}~\bibnamefont {Tong}}, \bibinfo {author}
  {\bibfnamefont {Y.}~\bibnamefont {Pang}},\ and\ \bibinfo {author}
  {\bibfnamefont {J.}~\bibnamefont {Xu}},\ }\href@noop {} {\bibfield  {journal}
  {\bibinfo  {journal} {Advanced Materials}\ }\textbf {\bibinfo {volume}
  {37}},\ \bibinfo {pages} {2400332} (\bibinfo {year} {2025})}\BibitemShut
  {NoStop}%
\bibitem [{\citenamefont {Park}\ \emph {et~al.}(2025)\citenamefont {Park},
  \citenamefont {Lee}, \citenamefont {Kim}, \citenamefont {Yoon}, \citenamefont
  {Park}, \citenamefont {Park}, \citenamefont {Song}, \citenamefont {Jeong},
  \citenamefont {Yoo},\ and\ \citenamefont {Yoo}}]{park2025brain}%
  \BibitemOpen
  \bibfield  {author} {\bibinfo {author} {\bibfnamefont {S.}~\bibnamefont
  {Park}}, \bibinfo {author} {\bibfnamefont {Y.}~\bibnamefont {Lee}}, \bibinfo
  {author} {\bibfnamefont {H.}~\bibnamefont {Kim}}, \bibinfo {author}
  {\bibfnamefont {T.}~\bibnamefont {Yoon}}, \bibinfo {author} {\bibfnamefont
  {Y.}~\bibnamefont {Park}}, \bibinfo {author} {\bibfnamefont {J.-U.}\
  \bibnamefont {Park}}, \bibinfo {author} {\bibfnamefont {Y.~M.}\ \bibnamefont
  {Song}}, \bibinfo {author} {\bibfnamefont {J.}~\bibnamefont {Jeong}},
  \bibinfo {author} {\bibfnamefont {H.}~\bibnamefont {Yoo}},\ and\ \bibinfo
  {author} {\bibfnamefont {H.}~\bibnamefont {Yoo}},\ }\href
  {https://doi.org/10.1038/s41467-025-57352-1} {\bibfield  {journal} {\bibinfo
  {journal} {Nature Communications}\ }\textbf {\bibinfo {volume} {16}},\
  \bibinfo {pages} {57352} (\bibinfo {year} {2025})}\BibitemShut {NoStop}%
\bibitem [{\citenamefont {Blouw}\ and\ \citenamefont
  {Eliasmith}(2020)}]{9053043}%
  \BibitemOpen
  \bibfield  {author} {\bibinfo {author} {\bibfnamefont {P.}~\bibnamefont
  {Blouw}}\ and\ \bibinfo {author} {\bibfnamefont {C.}~\bibnamefont
  {Eliasmith}},\ }in\ \href {https://doi.org/10.1109/ICASSP40776.2020.9053043}
  {\emph {\bibinfo {booktitle} {ICASSP 2020 - 2020 IEEE International
  Conference on Acoustics, Speech and Signal Processing (ICASSP)}}}\ (\bibinfo
  {year} {2020})\ pp.\ \bibinfo {pages} {8534--8538}\BibitemShut {NoStop}%
\bibitem [{\citenamefont {Bhatnagar}\ and\ \citenamefont
  {Kumar}(2025)}]{bhatnagar2025comprehensive}%
  \BibitemOpen
  \bibfield  {author} {\bibinfo {author} {\bibfnamefont {V.}~\bibnamefont
  {Bhatnagar}}\ and\ \bibinfo {author} {\bibfnamefont {A.}~\bibnamefont
  {Kumar}},\ }\href@noop {} {\bibfield  {journal} {\bibinfo  {journal} {Energy
  Storage}\ }\textbf {\bibinfo {volume} {7}},\ \bibinfo {pages} {e70272}
  (\bibinfo {year} {2025})}\BibitemShut {NoStop}%
\bibitem [{\citenamefont {Cui}\ \emph {et~al.}(2025)\citenamefont {Cui},
  \citenamefont {Xiao}, \citenamefont {Yang}, \citenamefont {Pei},
  \citenamefont {Ke}, \citenamefont {Fang}, \citenamefont {Qiao}, \citenamefont
  {Shi}, \citenamefont {Long}, \citenamefont {Xu} \emph
  {et~al.}}]{cui2025bioinspired}%
  \BibitemOpen
  \bibfield  {author} {\bibinfo {author} {\bibfnamefont {H.}~\bibnamefont
  {Cui}}, \bibinfo {author} {\bibfnamefont {Y.}~\bibnamefont {Xiao}}, \bibinfo
  {author} {\bibfnamefont {Y.}~\bibnamefont {Yang}}, \bibinfo {author}
  {\bibfnamefont {M.}~\bibnamefont {Pei}}, \bibinfo {author} {\bibfnamefont
  {S.}~\bibnamefont {Ke}}, \bibinfo {author} {\bibfnamefont {X.}~\bibnamefont
  {Fang}}, \bibinfo {author} {\bibfnamefont {L.}~\bibnamefont {Qiao}}, \bibinfo
  {author} {\bibfnamefont {K.}~\bibnamefont {Shi}}, \bibinfo {author}
  {\bibfnamefont {H.}~\bibnamefont {Long}}, \bibinfo {author} {\bibfnamefont
  {W.}~\bibnamefont {Xu}}, \emph {et~al.},\ }\href@noop {} {\bibfield
  {journal} {\bibinfo  {journal} {Nature Communications}\ }\textbf {\bibinfo
  {volume} {16}},\ \bibinfo {pages} {2263} (\bibinfo {year}
  {2025})}\BibitemShut {NoStop}%
\bibitem [{\citenamefont {Jaeger}(2001)}]{jaeger2001echo}%
  \BibitemOpen
  \bibfield  {author} {\bibinfo {author} {\bibfnamefont {H.}~\bibnamefont
  {Jaeger}},\ }\href@noop {} {\emph {\bibinfo {title} {The "echo state"
  approach to analysing and training recurrent neural networks—with an
  erratum note}}},\ \bibinfo {type} {Technical Report}\ \bibinfo {number}
  {148}\ (\bibinfo  {institution} {German National Research Center for
  Information Technology GMD},\ \bibinfo {address} {Bonn, Germany},\ \bibinfo
  {year} {2001})\BibitemShut {NoStop}%
\bibitem [{\citenamefont {Maass}\ \emph {et~al.}(2002)\citenamefont {Maass},
  \citenamefont {Natschl{\"a}ger},\ and\ \citenamefont
  {Markram}}]{maass2002real}%
  \BibitemOpen
  \bibfield  {author} {\bibinfo {author} {\bibfnamefont {W.}~\bibnamefont
  {Maass}}, \bibinfo {author} {\bibfnamefont {T.}~\bibnamefont
  {Natschl{\"a}ger}},\ and\ \bibinfo {author} {\bibfnamefont {H.}~\bibnamefont
  {Markram}},\ }\href@noop {} {\bibfield  {journal} {\bibinfo  {journal}
  {Neural computation}\ }\textbf {\bibinfo {volume} {14}},\ \bibinfo {pages}
  {2531} (\bibinfo {year} {2002})}\BibitemShut {NoStop}%
\bibitem [{\citenamefont {Nakajima}\ and\ \citenamefont
  {Fischer}(2024)}]{nakajima2024review}%
  \BibitemOpen
  \bibfield  {author} {\bibinfo {author} {\bibfnamefont {K.}~\bibnamefont
  {Nakajima}}\ and\ \bibinfo {author} {\bibfnamefont {I.}~\bibnamefont
  {Fischer}},\ }\href {https://doi.org/10.1038/s41467-024-45187-1} {\bibfield
  {journal} {\bibinfo  {journal} {Nature Communications}\ }\textbf {\bibinfo
  {volume} {15}},\ \bibinfo {pages} {45187} (\bibinfo {year}
  {2024})}\BibitemShut {NoStop}%
\bibitem [{\citenamefont {{Verstraeten, David and Schrauwen, Benjamin and
  D'Haene, Michiel and Stroobandt, Dirk}}(2006)}]{374553}%
  \BibitemOpen
  \bibfield  {author} {\bibinfo {author} {\bibnamefont {{Verstraeten, David and
  Schrauwen, Benjamin and D'Haene, Michiel and Stroobandt, Dirk}}},\ }in\
  \href{http://doi.org/1854/10283}{\emph{\bibinfo {booktitle}{Proceedings of
  the 2006 EPFL LATSIS Symposium}}}\ (\bibinfo {year}{2006})\ pp.\ \bibinfo
  {pages}{139--140}\BibitemShut {NoStop}%

\bibitem [{\citenamefont
  {te~Vrugt}(2024)}]{vrugt2024introductionreservoircomputing}%
  \BibitemOpen
  \bibfield  {author} {\bibinfo {author} {\bibfnamefont {M.}~\bibnamefont
  {te~Vrugt}},\ }\href {https://arxiv.org/abs/2412.13212} {\bibinfo {title} {An
  introduction to reservoir computing}} (\bibinfo {year} {2024}),\ \Eprint
  {https://arxiv.org/abs/2412.13212} {arXiv:2412.13212 [cs.ET]} \BibitemShut
  {NoStop}%
\bibitem [{\citenamefont {Du}\ \emph {et~al.}(2025)\citenamefont {Du},
  \citenamefont {Luo}, \citenamefont {Guo}, \citenamefont {Xiao}, \citenamefont
  {Yu},\ and\ \citenamefont {Wang}}]{PhysRevE.111.035303}%
  \BibitemOpen
  \bibfield  {author} {\bibinfo {author} {\bibfnamefont {Y.}~\bibnamefont
  {Du}}, \bibinfo {author} {\bibfnamefont {H.}~\bibnamefont {Luo}}, \bibinfo
  {author} {\bibfnamefont {J.}~\bibnamefont {Guo}}, \bibinfo {author}
  {\bibfnamefont {J.}~\bibnamefont {Xiao}}, \bibinfo {author} {\bibfnamefont
  {Y.}~\bibnamefont {Yu}},\ and\ \bibinfo {author} {\bibfnamefont
  {X.}~\bibnamefont {Wang}},\ }\href
  {https://doi.org/10.1103/PhysRevE.111.035303} {\bibfield  {journal} {\bibinfo
   {journal} {Phys. Rev. E}\ }\textbf {\bibinfo {volume} {111}},\ \bibinfo
  {pages} {035303} (\bibinfo {year} {2025})}\BibitemShut {NoStop}%
\bibitem [{\citenamefont {Wringe}\ \emph {et~al.}(2025)\citenamefont {Wringe},
  \citenamefont {Trefzer},\ and\ \citenamefont {Stepney}}]{Wringe_2025}%
  \BibitemOpen
  \bibfield  {author} {\bibinfo {author} {\bibfnamefont {C.}~\bibnamefont
  {Wringe}}, \bibinfo {author} {\bibfnamefont {M.}~\bibnamefont {Trefzer}},\
  and\ \bibinfo {author} {\bibfnamefont {S.}~\bibnamefont {Stepney}},\ }\href
  {https://doi.org/10.1080/17445760.2025.2472211} {\bibfield  {journal}
  {\bibinfo  {journal} {International Journal of Parallel, Emergent and
  Distributed Systems}\ ,\ \bibinfo {pages} {1–39}} (\bibinfo {year}
  {2025})}\BibitemShut {NoStop}%
\bibitem [{\citenamefont {Fujii}\ and\ \citenamefont
  {Nakajima}(2017)}]{Fujii_2017}%
  \BibitemOpen
  \bibfield  {author} {\bibinfo {author} {\bibfnamefont {K.}~\bibnamefont
  {Fujii}}\ and\ \bibinfo {author} {\bibfnamefont {K.}~\bibnamefont
  {Nakajima}},\ }\bibfield  {journal} {\bibinfo  {journal} {Physical Review
  Applied}\ }\textbf {\bibinfo {volume} {8}},\ \href
  {https://doi.org/10.1103/physrevapplied.8.024030}
  {10.1103/physrevapplied.8.024030} (\bibinfo {year} {2017})\BibitemShut
  {NoStop}%
\bibitem [{\citenamefont {Ghosh}\ \emph {et~al.}(2019)\citenamefont {Ghosh},
  \citenamefont {Opala}, \citenamefont {Matuszewski}, \citenamefont {Paterek},\
  and\ \citenamefont {Fazio}}]{ghosh2019quantum}%
  \BibitemOpen
  \bibfield  {author} {\bibinfo {author} {\bibfnamefont {S.}~\bibnamefont
  {Ghosh}}, \bibinfo {author} {\bibfnamefont {A.}~\bibnamefont {Opala}},
  \bibinfo {author} {\bibfnamefont {M.}~\bibnamefont {Matuszewski}}, \bibinfo
  {author} {\bibfnamefont {T.}~\bibnamefont {Paterek}},\ and\ \bibinfo {author}
  {\bibfnamefont {R.}~\bibnamefont {Fazio}},\ }\href
  {https://doi.org/10.1038/s41534-019-0149-8} {\bibfield  {journal} {\bibinfo
  {journal} {npj Quantum Information}\ }\textbf {\bibinfo {volume} {5}},\
  \bibinfo {pages} {35} (\bibinfo {year} {2019})}\BibitemShut {NoStop}%
\bibitem [{\citenamefont {Chen}\ \emph {et~al.}(2020)\citenamefont {Chen},
  \citenamefont {Nurdin},\ and\ \citenamefont
  {Yamamoto}}]{PhysRevApplied.14.024065}%
  \BibitemOpen
  \bibfield  {author} {\bibinfo {author} {\bibfnamefont {J.}~\bibnamefont
  {Chen}}, \bibinfo {author} {\bibfnamefont {H.~I.}\ \bibnamefont {Nurdin}},\
  and\ \bibinfo {author} {\bibfnamefont {N.}~\bibnamefont {Yamamoto}},\ }\href
  {https://doi.org/10.1103/PhysRevApplied.14.024065} {\bibfield  {journal}
  {\bibinfo  {journal} {Phys. Rev. Appl.}\ }\textbf {\bibinfo {volume} {14}},\
  \bibinfo {pages} {024065} (\bibinfo {year} {2020})}\BibitemShut {NoStop}%
\bibitem [{\citenamefont {Gyurik}\ \emph {et~al.}(2025)\citenamefont {Gyurik},
  \citenamefont {Wudarski}, \citenamefont {Philip}, \citenamefont {Sannia},
  \citenamefont {Sadeghi}, \citenamefont {Kyriienko}, \citenamefont
  {Venturelli},\ and\ \citenamefont
  {Gentile}}]{gyurik2025quantumfeaturemapsquantum}%
  \BibitemOpen
  \bibfield  {author} {\bibinfo {author} {\bibfnamefont {C.}~\bibnamefont
  {Gyurik}}, \bibinfo {author} {\bibfnamefont {F.}~\bibnamefont {Wudarski}},
  \bibinfo {author} {\bibfnamefont {E.}~\bibnamefont {Philip}}, \bibinfo
  {author} {\bibfnamefont {A.}~\bibnamefont {Sannia}}, \bibinfo {author}
  {\bibfnamefont {H.}~\bibnamefont {Sadeghi}}, \bibinfo {author} {\bibfnamefont
  {O.}~\bibnamefont {Kyriienko}}, \bibinfo {author} {\bibfnamefont
  {D.}~\bibnamefont {Venturelli}},\ and\ \bibinfo {author} {\bibfnamefont
  {A.~A.}\ \bibnamefont {Gentile}},\ }\href {https://arxiv.org/abs/2510.01797}
  {\bibinfo {title} {From quantum feature maps to quantum reservoir computing:
  perspectives and applications}} (\bibinfo {year} {2025}),\ \Eprint
  {https://arxiv.org/abs/2510.01797} {arXiv:2510.01797 [quant-ph]} \BibitemShut
  {NoStop}%
\bibitem [{\citenamefont {Pfeffer}\ \emph {et~al.}(2022)\citenamefont
  {Pfeffer}, \citenamefont {Heyder},\ and\ \citenamefont
  {Schumacher}}]{PhysRevResearch.4.033176}%
  \BibitemOpen
  \bibfield  {author} {\bibinfo {author} {\bibfnamefont {P.}~\bibnamefont
  {Pfeffer}}, \bibinfo {author} {\bibfnamefont {F.}~\bibnamefont {Heyder}},\
  and\ \bibinfo {author} {\bibfnamefont {J.}~\bibnamefont {Schumacher}},\
  }\href {https://doi.org/10.1103/PhysRevResearch.4.033176} {\bibfield
  {journal} {\bibinfo  {journal} {Phys. Rev. Res.}\ }\textbf {\bibinfo {volume}
  {4}},\ \bibinfo {pages} {033176} (\bibinfo {year} {2022})}\BibitemShut
  {NoStop}%
\bibitem [{\citenamefont {Kora}\ and\ \citenamefont
  {Simon}(2025)}]{kora2025statistical}%
  \BibitemOpen
  \bibfield  {author} {\bibinfo {author} {\bibfnamefont {Y.}~\bibnamefont
  {Kora}}\ and\ \bibinfo {author} {\bibfnamefont {C.}~\bibnamefont {Simon}},\
  }\href@noop {} {\bibfield  {journal} {\bibinfo  {journal} {arXiv preprint
  arXiv:2504.17837}\ } (\bibinfo {year} {2025})}\BibitemShut {NoStop}%
\bibitem [{\citenamefont {Kora}\ \emph {et~al.}(2024)\citenamefont {Kora},
  \citenamefont {Zadeh-Haghighi}, \citenamefont {Stewart}, \citenamefont
  {Heshami},\ and\ \citenamefont
  {Simon}}]{kora2024frequencydissipationdependententanglementadvantage}%
  \BibitemOpen
  \bibfield  {author} {\bibinfo {author} {\bibfnamefont {Y.}~\bibnamefont
  {Kora}}, \bibinfo {author} {\bibfnamefont {H.}~\bibnamefont
  {Zadeh-Haghighi}}, \bibinfo {author} {\bibfnamefont {T.~C.}\ \bibnamefont
  {Stewart}}, \bibinfo {author} {\bibfnamefont {K.}~\bibnamefont {Heshami}},\
  and\ \bibinfo {author} {\bibfnamefont {C.}~\bibnamefont {Simon}},\ }\href
  {https://arxiv.org/abs/2403.08998} {\bibinfo {title} {Frequency- and
  dissipation-dependent entanglement advantage in spin-network quantum
  reservoir computing}} (\bibinfo {year} {2024}),\ \Eprint
  {https://arxiv.org/abs/2403.08998} {arXiv:2403.08998 [quant-ph]} \BibitemShut
  {NoStop}%
\bibitem [{\citenamefont {G\"otting}\ \emph {et~al.}(2023)\citenamefont
  {G\"otting}, \citenamefont {Lohof},\ and\ \citenamefont
  {Gies}}]{PhysRevA.108.052427}%
  \BibitemOpen
  \bibfield  {author} {\bibinfo {author} {\bibfnamefont {N.}~\bibnamefont
  {G\"otting}}, \bibinfo {author} {\bibfnamefont {F.}~\bibnamefont {Lohof}},\
  and\ \bibinfo {author} {\bibfnamefont {C.}~\bibnamefont {Gies}},\ }\href
  {https://doi.org/10.1103/PhysRevA.108.052427} {\bibfield  {journal} {\bibinfo
   {journal} {Phys. Rev. A}\ }\textbf {\bibinfo {volume} {108}},\ \bibinfo
  {pages} {052427} (\bibinfo {year} {2023})}\BibitemShut {NoStop}%
\bibitem [{\citenamefont {Settino}\ \emph {et~al.}(2024)\citenamefont
  {Settino}, \citenamefont {Salatino}, \citenamefont {Mariani}, \citenamefont
  {Channab}, \citenamefont {Bozzolo}, \citenamefont {Vallisa}, \citenamefont
  {Barill{\`a}}, \citenamefont {Policicchio}, \citenamefont {Gullo},
  \citenamefont {Giordano} \emph {et~al.}}]{settino2024memory}%
  \BibitemOpen
  \bibfield  {author} {\bibinfo {author} {\bibfnamefont {J.}~\bibnamefont
  {Settino}}, \bibinfo {author} {\bibfnamefont {L.}~\bibnamefont {Salatino}},
  \bibinfo {author} {\bibfnamefont {L.}~\bibnamefont {Mariani}}, \bibinfo
  {author} {\bibfnamefont {M.}~\bibnamefont {Channab}}, \bibinfo {author}
  {\bibfnamefont {L.}~\bibnamefont {Bozzolo}}, \bibinfo {author} {\bibfnamefont
  {S.}~\bibnamefont {Vallisa}}, \bibinfo {author} {\bibfnamefont
  {P.}~\bibnamefont {Barill{\`a}}}, \bibinfo {author} {\bibfnamefont
  {A.}~\bibnamefont {Policicchio}}, \bibinfo {author} {\bibfnamefont {N.~L.}\
  \bibnamefont {Gullo}}, \bibinfo {author} {\bibfnamefont {A.}~\bibnamefont
  {Giordano}}, \emph {et~al.},\ }\href@noop {} {\bibfield  {journal} {\bibinfo
  {journal} {arXiv e-prints}\ ,\ \bibinfo {pages} {arXiv}} (\bibinfo {year}
  {2024})}\BibitemShut {NoStop}%
\bibitem [{\citenamefont {Kornjača}\ \emph {et~al.}(2024)\citenamefont
  {Kornjača}, \citenamefont {Hu}, \citenamefont {Zhao}, \citenamefont {Wurtz},
  \citenamefont {Weinberg}, \citenamefont {Hamdan}, \citenamefont {Zhdanov},
  \citenamefont {Cantu}, \citenamefont {Zhou}, \citenamefont {Bravo},
  \citenamefont {Bagnall}, \citenamefont {Basham}, \citenamefont {Campo},
  \citenamefont {Choukri}, \citenamefont {DeAngelo}, \citenamefont {Frederick},
  \citenamefont {Haines}, \citenamefont {Hammett}, \citenamefont {Hsu},
  \citenamefont {Hu}, \citenamefont {Huber}, \citenamefont {Jepsen},
  \citenamefont {Jia}, \citenamefont {Karolyshyn}, \citenamefont {Kwon},
  \citenamefont {Long}, \citenamefont {Lopatin}, \citenamefont {Lukin},
  \citenamefont {Macrì}, \citenamefont {Marković}, \citenamefont
  {Martínez-Martínez}, \citenamefont {Meng}, \citenamefont {Ostroumov},
  \citenamefont {Paquette}, \citenamefont {Robinson}, \citenamefont
  {Rodriguez}, \citenamefont {Singh}, \citenamefont {Sinha}, \citenamefont
  {Thoreen}, \citenamefont {Wan}, \citenamefont {Waxman-Lenz}, \citenamefont
  {Wong}, \citenamefont {Wu}, \citenamefont {Lopes}, \citenamefont {Boger},
  \citenamefont {Gemelke}, \citenamefont {Kitagawa}, \citenamefont {Keesling},
  \citenamefont {Gao}, \citenamefont {Bylinskii}, \citenamefont {Yelin},
  \citenamefont {Liu},\ and\ \citenamefont
  {Wang}}]{kornjača2024largescalequantumreservoirlearning}%
  \BibitemOpen
  \bibfield  {author} {\bibinfo {author} {\bibfnamefont {M.}~\bibnamefont
  {Kornjača}}, \bibinfo {author} {\bibfnamefont {H.-Y.}\ \bibnamefont {Hu}},
  \bibinfo {author} {\bibfnamefont {C.}~\bibnamefont {Zhao}}, \bibinfo {author}
  {\bibfnamefont {J.}~\bibnamefont {Wurtz}}, \bibinfo {author} {\bibfnamefont
  {P.}~\bibnamefont {Weinberg}}, \bibinfo {author} {\bibfnamefont
  {M.}~\bibnamefont {Hamdan}}, \bibinfo {author} {\bibfnamefont
  {A.}~\bibnamefont {Zhdanov}}, \bibinfo {author} {\bibfnamefont {S.~H.}\
  \bibnamefont {Cantu}}, \bibinfo {author} {\bibfnamefont {H.}~\bibnamefont
  {Zhou}}, \bibinfo {author} {\bibfnamefont {R.~A.}\ \bibnamefont {Bravo}},
  \bibinfo {author} {\bibfnamefont {K.}~\bibnamefont {Bagnall}}, \bibinfo
  {author} {\bibfnamefont {J.~I.}\ \bibnamefont {Basham}}, \bibinfo {author}
  {\bibfnamefont {J.}~\bibnamefont {Campo}}, \bibinfo {author} {\bibfnamefont
  {A.}~\bibnamefont {Choukri}}, \bibinfo {author} {\bibfnamefont
  {R.}~\bibnamefont {DeAngelo}}, \bibinfo {author} {\bibfnamefont
  {P.}~\bibnamefont {Frederick}}, \bibinfo {author} {\bibfnamefont
  {D.}~\bibnamefont {Haines}}, \bibinfo {author} {\bibfnamefont
  {J.}~\bibnamefont {Hammett}}, \bibinfo {author} {\bibfnamefont
  {N.}~\bibnamefont {Hsu}}, \bibinfo {author} {\bibfnamefont {M.-G.}\
  \bibnamefont {Hu}}, \bibinfo {author} {\bibfnamefont {F.}~\bibnamefont
  {Huber}}, \bibinfo {author} {\bibfnamefont {P.~N.}\ \bibnamefont {Jepsen}},
  \bibinfo {author} {\bibfnamefont {N.}~\bibnamefont {Jia}}, \bibinfo {author}
  {\bibfnamefont {T.}~\bibnamefont {Karolyshyn}}, \bibinfo {author}
  {\bibfnamefont {M.}~\bibnamefont {Kwon}}, \bibinfo {author} {\bibfnamefont
  {J.}~\bibnamefont {Long}}, \bibinfo {author} {\bibfnamefont {J.}~\bibnamefont
  {Lopatin}}, \bibinfo {author} {\bibfnamefont {A.}~\bibnamefont {Lukin}},
  \bibinfo {author} {\bibfnamefont {T.}~\bibnamefont {Macrì}}, \bibinfo
  {author} {\bibfnamefont {O.}~\bibnamefont {Marković}}, \bibinfo {author}
  {\bibfnamefont {L.~A.}\ \bibnamefont {Martínez-Martínez}}, \bibinfo
  {author} {\bibfnamefont {X.}~\bibnamefont {Meng}}, \bibinfo {author}
  {\bibfnamefont {E.}~\bibnamefont {Ostroumov}}, \bibinfo {author}
  {\bibfnamefont {D.}~\bibnamefont {Paquette}}, \bibinfo {author}
  {\bibfnamefont {J.}~\bibnamefont {Robinson}}, \bibinfo {author}
  {\bibfnamefont {P.~S.}\ \bibnamefont {Rodriguez}}, \bibinfo {author}
  {\bibfnamefont {A.}~\bibnamefont {Singh}}, \bibinfo {author} {\bibfnamefont
  {N.}~\bibnamefont {Sinha}}, \bibinfo {author} {\bibfnamefont
  {H.}~\bibnamefont {Thoreen}}, \bibinfo {author} {\bibfnamefont
  {N.}~\bibnamefont {Wan}}, \bibinfo {author} {\bibfnamefont {D.}~\bibnamefont
  {Waxman-Lenz}}, \bibinfo {author} {\bibfnamefont {T.}~\bibnamefont {Wong}},
  \bibinfo {author} {\bibfnamefont {K.-H.}\ \bibnamefont {Wu}}, \bibinfo
  {author} {\bibfnamefont {P.~L.~S.}\ \bibnamefont {Lopes}}, \bibinfo {author}
  {\bibfnamefont {Y.}~\bibnamefont {Boger}}, \bibinfo {author} {\bibfnamefont
  {N.}~\bibnamefont {Gemelke}}, \bibinfo {author} {\bibfnamefont
  {T.}~\bibnamefont {Kitagawa}}, \bibinfo {author} {\bibfnamefont
  {A.}~\bibnamefont {Keesling}}, \bibinfo {author} {\bibfnamefont
  {X.}~\bibnamefont {Gao}}, \bibinfo {author} {\bibfnamefont {A.}~\bibnamefont
  {Bylinskii}}, \bibinfo {author} {\bibfnamefont {S.~F.}\ \bibnamefont
  {Yelin}}, \bibinfo {author} {\bibfnamefont {F.}~\bibnamefont {Liu}},\ and\
  \bibinfo {author} {\bibfnamefont {S.-T.}\ \bibnamefont {Wang}},\ }\href
  {https://arxiv.org/abs/2407.02553} {\bibinfo {title} {Large-scale quantum
  reservoir learning with an analog quantum computer}} (\bibinfo {year}
  {2024}),\ \Eprint {https://arxiv.org/abs/2407.02553} {arXiv:2407.02553
  [quant-ph]} \BibitemShut {NoStop}%
\bibitem [{\citenamefont {Bravo}\ \emph {et~al.}(2022)\citenamefont {Bravo},
  \citenamefont {Najafi}, \citenamefont {Gao},\ and\ \citenamefont
  {Yelin}}]{PRXQuantum.3.030325}%
  \BibitemOpen
  \bibfield  {author} {\bibinfo {author} {\bibfnamefont {R.~A.}\ \bibnamefont
  {Bravo}}, \bibinfo {author} {\bibfnamefont {K.}~\bibnamefont {Najafi}},
  \bibinfo {author} {\bibfnamefont {X.}~\bibnamefont {Gao}},\ and\ \bibinfo
  {author} {\bibfnamefont {S.~F.}\ \bibnamefont {Yelin}},\ }\href
  {https://doi.org/10.1103/PRXQuantum.3.030325} {\bibfield  {journal} {\bibinfo
   {journal} {PRX Quantum}\ }\textbf {\bibinfo {volume} {3}},\ \bibinfo {pages}
  {030325} (\bibinfo {year} {2022})}\BibitemShut {NoStop}%
\bibitem [{\citenamefont {Beaulieu}\ \emph {et~al.}(2025)\citenamefont
  {Beaulieu}, \citenamefont {Kornjača}, \citenamefont {Krunic}, \citenamefont
  {Stivaktakis}, \citenamefont {Chen}, \citenamefont {Ehmer}, \citenamefont
  {Wang},\ and\ \citenamefont {Pham}}]{beaulieu2025robust}%
  \BibitemOpen
  \bibfield  {author} {\bibinfo {author} {\bibfnamefont {D.}~\bibnamefont
  {Beaulieu}}, \bibinfo {author} {\bibfnamefont {M.}~\bibnamefont {Kornjača}},
  \bibinfo {author} {\bibfnamefont {Z.}~\bibnamefont {Krunic}}, \bibinfo
  {author} {\bibfnamefont {M.}~\bibnamefont {Stivaktakis}}, \bibinfo {author}
  {\bibfnamefont {J.}~\bibnamefont {Chen}}, \bibinfo {author} {\bibfnamefont
  {T.}~\bibnamefont {Ehmer}}, \bibinfo {author} {\bibfnamefont {S.-T.}\
  \bibnamefont {Wang}},\ and\ \bibinfo {author} {\bibfnamefont
  {A.}~\bibnamefont {Pham}},\ }\href@noop {} {\bibfield  {journal} {\bibinfo
  {journal} {Journal of Chemical Information and Modeling}\ }\textbf {\bibinfo
  {volume} {65}},\ \bibinfo {pages} {8475} (\bibinfo {year}
  {2025})}\BibitemShut {NoStop}%
\bibitem [{\citenamefont {Llodr{\`a}}\ \emph {et~al.}(2025)\citenamefont
  {Llodr{\`a}}, \citenamefont {Mujal}, \citenamefont {Zambrini},\ and\
  \citenamefont {Giorgi}}]{llodra2025quantum}%
  \BibitemOpen
  \bibfield  {author} {\bibinfo {author} {\bibfnamefont {G.}~\bibnamefont
  {Llodr{\`a}}}, \bibinfo {author} {\bibfnamefont {P.}~\bibnamefont {Mujal}},
  \bibinfo {author} {\bibfnamefont {R.}~\bibnamefont {Zambrini}},\ and\
  \bibinfo {author} {\bibfnamefont {G.~L.}\ \bibnamefont {Giorgi}},\
  }\href@noop {} {\bibfield  {journal} {\bibinfo  {journal} {Chaos, Solitons \&
  Fractals}\ }\textbf {\bibinfo {volume} {195}},\ \bibinfo {pages} {116289}
  (\bibinfo {year} {2025})}\BibitemShut {NoStop}%
\bibitem [{\citenamefont {Govia}\ \emph {et~al.}(2021)\citenamefont {Govia},
  \citenamefont {Ribeill}, \citenamefont {Rowlands}, \citenamefont {Krovi},\
  and\ \citenamefont {Ohki}}]{PhysRevResearch.3.013077}%
  \BibitemOpen
  \bibfield  {author} {\bibinfo {author} {\bibfnamefont {L.~C.~G.}\
  \bibnamefont {Govia}}, \bibinfo {author} {\bibfnamefont {G.~J.}\ \bibnamefont
  {Ribeill}}, \bibinfo {author} {\bibfnamefont {G.~E.}\ \bibnamefont
  {Rowlands}}, \bibinfo {author} {\bibfnamefont {H.~K.}\ \bibnamefont
  {Krovi}},\ and\ \bibinfo {author} {\bibfnamefont {T.~A.}\ \bibnamefont
  {Ohki}},\ }\href {https://doi.org/10.1103/PhysRevResearch.3.013077}
  {\bibfield  {journal} {\bibinfo  {journal} {Phys. Rev. Res.}\ }\textbf
  {\bibinfo {volume} {3}},\ \bibinfo {pages} {013077} (\bibinfo {year}
  {2021})}\BibitemShut {NoStop}%
\bibitem [{\citenamefont {Karimi}\ \emph {et~al.}(2025)\citenamefont {Karimi},
  \citenamefont {Zadeh-Haghighi}, \citenamefont {Kora},\ and\ \citenamefont
  {Simon}}]{karimi2025role}%
  \BibitemOpen
  \bibfield  {author} {\bibinfo {author} {\bibfnamefont {A.}~\bibnamefont
  {Karimi}}, \bibinfo {author} {\bibfnamefont {H.}~\bibnamefont
  {Zadeh-Haghighi}}, \bibinfo {author} {\bibfnamefont {Y.}~\bibnamefont
  {Kora}},\ and\ \bibinfo {author} {\bibfnamefont {C.}~\bibnamefont {Simon}},\
  }\href@noop {} {\bibfield  {journal} {\bibinfo  {journal} {arXiv preprint
  arXiv:2508.11175}\ } (\bibinfo {year} {2025})}\BibitemShut {NoStop}%
\bibitem [{\citenamefont {Hamhoum}\ \emph {et~al.}(2025)\citenamefont
  {Hamhoum}, \citenamefont {Cherkaoui}, \citenamefont {Laprade}, \citenamefont
  {Ahmed},\ and\ \citenamefont
  {Wang}}]{hamhoum2025multivariatetimeseriesforecasting}%
  \BibitemOpen
  \bibfield  {author} {\bibinfo {author} {\bibfnamefont {W.}~\bibnamefont
  {Hamhoum}}, \bibinfo {author} {\bibfnamefont {S.}~\bibnamefont {Cherkaoui}},
  \bibinfo {author} {\bibfnamefont {J.-F.}\ \bibnamefont {Laprade}}, \bibinfo
  {author} {\bibfnamefont {O.}~\bibnamefont {Ahmed}},\ and\ \bibinfo {author}
  {\bibfnamefont {S.}~\bibnamefont {Wang}},\ }\href
  {https://arxiv.org/abs/2510.13634} {\bibinfo {title} {Multivariate time
  series forecasting with gate-based quantum reservoir computing on nisq
  hardware}} (\bibinfo {year} {2025}),\ \Eprint
  {https://arxiv.org/abs/2510.13634} {arXiv:2510.13634 [cs.LG]} \BibitemShut
  {NoStop}%
\bibitem [{\citenamefont {Li}\ \emph {et~al.}(2025)\citenamefont {Li},
  \citenamefont {Mukhopadhyay}, \citenamefont {Bayat},\ and\ \citenamefont
  {Habibnia}}]{li2025quantumreservoircomputingrealized}%
  \BibitemOpen
  \bibfield  {author} {\bibinfo {author} {\bibfnamefont {Q.}~\bibnamefont
  {Li}}, \bibinfo {author} {\bibfnamefont {C.}~\bibnamefont {Mukhopadhyay}},
  \bibinfo {author} {\bibfnamefont {A.}~\bibnamefont {Bayat}},\ and\ \bibinfo
  {author} {\bibfnamefont {A.}~\bibnamefont {Habibnia}},\ }\href
  {https://arxiv.org/abs/2505.13933} {\bibinfo {title} {Quantum reservoir
  computing for realized volatility forecasting}} (\bibinfo {year} {2025}),\
  \Eprint {https://arxiv.org/abs/2505.13933} {arXiv:2505.13933 [quant-ph]}
  \BibitemShut {NoStop}%
\bibitem [{\citenamefont {Gilboa}\ \emph {et~al.}(2024)\citenamefont {Gilboa},
  \citenamefont {Michaeli}, \citenamefont {Soudry},\ and\ \citenamefont
  {McClean}}]{gilboa2024exponentialquantumcommunicationadvantage}%
  \BibitemOpen
  \bibfield  {author} {\bibinfo {author} {\bibfnamefont {D.}~\bibnamefont
  {Gilboa}}, \bibinfo {author} {\bibfnamefont {H.}~\bibnamefont {Michaeli}},
  \bibinfo {author} {\bibfnamefont {D.}~\bibnamefont {Soudry}},\ and\ \bibinfo
  {author} {\bibfnamefont {J.~R.}\ \bibnamefont {McClean}},\ }\href
  {https://arxiv.org/abs/2310.07136} {\bibinfo {title} {Exponential quantum
  communication advantage in distributed inference and learning}} (\bibinfo
  {year} {2024}),\ \Eprint {https://arxiv.org/abs/2310.07136} {arXiv:2310.07136
  [quant-ph]} \BibitemShut {NoStop}%
\bibitem [{\citenamefont {Ho}\ \emph {et~al.}(2022)\citenamefont {Ho},
  \citenamefont {Moreno}, \citenamefont {Brito}, \citenamefont {Graffitti},
  \citenamefont {Morrison}, \citenamefont {Nery}, \citenamefont {Pickston},
  \citenamefont {Proietti}, \citenamefont {Rabelo}, \citenamefont {Fedrizzi}
  \emph {et~al.}}]{ho2022entanglement}%
  \BibitemOpen
  \bibfield  {author} {\bibinfo {author} {\bibfnamefont {J.}~\bibnamefont
  {Ho}}, \bibinfo {author} {\bibfnamefont {G.}~\bibnamefont {Moreno}}, \bibinfo
  {author} {\bibfnamefont {S.}~\bibnamefont {Brito}}, \bibinfo {author}
  {\bibfnamefont {F.}~\bibnamefont {Graffitti}}, \bibinfo {author}
  {\bibfnamefont {C.~L.}\ \bibnamefont {Morrison}}, \bibinfo {author}
  {\bibfnamefont {R.}~\bibnamefont {Nery}}, \bibinfo {author} {\bibfnamefont
  {A.}~\bibnamefont {Pickston}}, \bibinfo {author} {\bibfnamefont
  {M.}~\bibnamefont {Proietti}}, \bibinfo {author} {\bibfnamefont
  {R.}~\bibnamefont {Rabelo}}, \bibinfo {author} {\bibfnamefont
  {A.}~\bibnamefont {Fedrizzi}}, \emph {et~al.},\ }\href@noop {} {\bibfield
  {journal} {\bibinfo  {journal} {npj Quantum Information}\ }\textbf {\bibinfo
  {volume} {8}},\ \bibinfo {pages} {13} (\bibinfo {year} {2022})}\BibitemShut
  {NoStop}%
\bibitem [{\citenamefont {Hasegawa}\ \emph {et~al.}(2025)\citenamefont
  {Hasegawa}, \citenamefont {Gall},\ and\ \citenamefont
  {Modanese}}]{hasegawa2025maximumseparationquantumcommunication}%
  \BibitemOpen
  \bibfield  {author} {\bibinfo {author} {\bibfnamefont {A.}~\bibnamefont
  {Hasegawa}}, \bibinfo {author} {\bibfnamefont {F.~L.}\ \bibnamefont {Gall}},\
  and\ \bibinfo {author} {\bibfnamefont {A.}~\bibnamefont {Modanese}},\ }\href
  {https://arxiv.org/abs/2505.16457} {\bibinfo {title} {Maximum separation of
  quantum communication complexity with and without shared entanglement}}
  (\bibinfo {year} {2025}),\ \Eprint {https://arxiv.org/abs/2505.16457}
  {arXiv:2505.16457 [quant-ph]} \BibitemShut {NoStop}%
\bibitem [{\citenamefont {Jaeger}(2002{\natexlab{a}})}]{jaeger2002shortterm}%
  \BibitemOpen
  \bibfield  {author} {\bibinfo {author} {\bibfnamefont {H.}~\bibnamefont
  {Jaeger}},\ }\href {https://www.nature.com/articles/srep00514} {\bibfield
  {journal} {\bibinfo  {journal} {GMD Report 152, German National Research
  Center for Information Technology}\ } (\bibinfo {year}
  {2002}{\natexlab{a}})}\BibitemShut {NoStop}%
\bibitem [{\citenamefont {Chen}\ and\ \citenamefont
  {Kuo}(2025)}]{chen2025unravelingquantumenvironmentstransformerassisted}%
  \BibitemOpen
  \bibfield  {author} {\bibinfo {author} {\bibfnamefont {C.-S.}\ \bibnamefont
  {Chen}}\ and\ \bibinfo {author} {\bibfnamefont {E.-J.}\ \bibnamefont {Kuo}},\
  }\href {https://arxiv.org/abs/2505.06928} {\bibinfo {title} {Unraveling
  quantum environments: Transformer-assisted learning in lindblad dynamics}}
  (\bibinfo {year} {2025}),\ \Eprint {https://arxiv.org/abs/2505.06928}
  {arXiv:2505.06928 [quant-ph]} \BibitemShut {NoStop}%
\bibitem [{\citenamefont {Mart{\'i}nez-Pe{\~n}a}\ \emph
  {et~al.}(2023)\citenamefont {Mart{\'i}nez-Pe{\~n}a}, \citenamefont {Nokkala},
  \citenamefont {Giorgi}, \citenamefont {Zambrini},\ and\ \citenamefont
  {Soriano}}]{MartinezPena2023}%
  \BibitemOpen
  \bibfield  {author} {\bibinfo {author} {\bibfnamefont {R.}~\bibnamefont
  {Mart{\'i}nez-Pe{\~n}a}}, \bibinfo {author} {\bibfnamefont {J.}~\bibnamefont
  {Nokkala}}, \bibinfo {author} {\bibfnamefont {G.~L.}\ \bibnamefont {Giorgi}},
  \bibinfo {author} {\bibfnamefont {R.}~\bibnamefont {Zambrini}},\ and\
  \bibinfo {author} {\bibfnamefont {M.~C.}\ \bibnamefont {Soriano}},\ }\href
  {https://doi.org/10.1007/s12559-020-09772-y} {\bibfield  {journal} {\bibinfo
  {journal} {Cognitive Computation}\ }\textbf {\bibinfo {volume} {15}},\
  \bibinfo {pages} {1440} (\bibinfo {year} {2023})}\BibitemShut {NoStop}%
\bibitem [{\citenamefont {Wang}\ \emph {et~al.}(2023)\citenamefont {Wang},
  \citenamefont {Mulvihill}, \citenamefont {Hu}, \citenamefont {Lyu},
  \citenamefont {Shivpuje}, \citenamefont {Liu}, \citenamefont {Soley},
  \citenamefont {Geva}, \citenamefont {Batista},\ and\ \citenamefont
  {Kais}}]{Wang2023}%
  \BibitemOpen
  \bibfield  {author} {\bibinfo {author} {\bibfnamefont {Y.}~\bibnamefont
  {Wang}}, \bibinfo {author} {\bibfnamefont {E.}~\bibnamefont {Mulvihill}},
  \bibinfo {author} {\bibfnamefont {Z.}~\bibnamefont {Hu}}, \bibinfo {author}
  {\bibfnamefont {N.}~\bibnamefont {Lyu}}, \bibinfo {author} {\bibfnamefont
  {S.}~\bibnamefont {Shivpuje}}, \bibinfo {author} {\bibfnamefont
  {Y.}~\bibnamefont {Liu}}, \bibinfo {author} {\bibfnamefont {M.~B.}\
  \bibnamefont {Soley}}, \bibinfo {author} {\bibfnamefont {E.}~\bibnamefont
  {Geva}}, \bibinfo {author} {\bibfnamefont {V.~S.}\ \bibnamefont {Batista}},\
  and\ \bibinfo {author} {\bibfnamefont {S.}~\bibnamefont {Kais}},\ }\href
  {https://doi.org/10.1021/acs.jctc.3c00316} {\bibfield  {journal} {\bibinfo
  {journal} {Journal of Chemical Theory and Computation}\ }\textbf {\bibinfo
  {volume} {19}},\ \bibinfo {pages} {4851} (\bibinfo {year}
  {2023})}\BibitemShut {NoStop}%
\bibitem [{\citenamefont {Miramont}\ \emph {et~al.}(2024)\citenamefont
  {Miramont}, \citenamefont {Bardenet}, \citenamefont {Chainais},\ and\
  \citenamefont
  {Auger}}]{miramont2024benchmarkingmulticomponentsignalprocessing}%
  \BibitemOpen
  \bibfield  {author} {\bibinfo {author} {\bibfnamefont {J.~M.}\ \bibnamefont
  {Miramont}}, \bibinfo {author} {\bibfnamefont {R.}~\bibnamefont {Bardenet}},
  \bibinfo {author} {\bibfnamefont {P.}~\bibnamefont {Chainais}},\ and\
  \bibinfo {author} {\bibfnamefont {F.}~\bibnamefont {Auger}},\ }\href
  {https://arxiv.org/abs/2402.08521} {\bibinfo {title} {Benchmarking
  multi-component signal processing methods in the time-frequency plane}}
  (\bibinfo {year} {2024}),\ \Eprint {https://arxiv.org/abs/2402.08521}
  {arXiv:2402.08521 [eess.SP]} \BibitemShut {NoStop}%
\bibitem [{\citenamefont {Griffith}(2021)}]{Griffith2021}%
  \BibitemOpen
  \bibfield  {author} {\bibinfo {author} {\bibfnamefont {A.}~\bibnamefont
  {Griffith}},\ }\emph {\bibinfo {title} {Essential Reservoir Computing}},\
  \href
  {https://ezproxy.lib.ucalgary.ca/login?qurl=https%3A%2F%2Fwww.proquest.com%2Fdissertations-theses%2Fessential-reservoir-computing%2Fdocview%2F2690997954%2Fse-2%3Faccountid%3D9838}
  {Ph.D. thesis},\ \bibinfo  {school} {ProQuest Dissertations \& Theses Global}
  (\bibinfo {year} {2021}),\ \bibinfo {note} {order No. 29310824}\BibitemShut
  {NoStop}%
\bibitem [{\citenamefont {Jaeger}(2002{\natexlab{b}})}]{article}%
  \BibitemOpen
  \bibfield  {author} {\bibinfo {author} {\bibfnamefont {H.}~\bibnamefont
  {Jaeger}},\ }\href@noop {} {\  (\bibinfo {year}
  {2002}{\natexlab{b}})}\BibitemShut {NoStop}%
\bibitem [{\citenamefont {Carroll}(2022)}]{Carroll2022OptimizingMI}%
  \BibitemOpen
  \bibfield  {author} {\bibinfo {author} {\bibfnamefont {T.~L.}\ \bibnamefont
  {Carroll}},\ }\href {https://api.semanticscholar.org/CorpusID:245704304}
  {\bibfield  {journal} {\bibinfo  {journal} {Chaos}\ }\textbf {\bibinfo
  {volume} {32 2}},\ \bibinfo {pages} {023123} (\bibinfo {year}
  {2022})}\BibitemShut {NoStop}%
\bibitem [{\citenamefont {Vidal}\ and\ \citenamefont
  {Werner}(2002)}]{Vidal_2002}%
  \BibitemOpen
  \bibfield  {author} {\bibinfo {author} {\bibfnamefont {G.}~\bibnamefont
  {Vidal}}\ and\ \bibinfo {author} {\bibfnamefont {R.~F.}\ \bibnamefont
  {Werner}},\ }\bibfield  {journal} {\bibinfo  {journal} {Physical Review A}\
  }\textbf {\bibinfo {volume} {65}},\ \href
  {https://doi.org/10.1103/physreva.65.032314} {10.1103/physreva.65.032314}
  (\bibinfo {year} {2002})\BibitemShut {NoStop}%
\end{thebibliography}

\end{document}